\documentclass{aa}
\usepackage{graphicx}
\usepackage{txfonts}
\usepackage{comment}
\usepackage{adjustbox}

\usepackage[breaklinks=true, colorlinks=true, citecolor=blue, linkcolor=blue,urlcolor=blue]{hyperref}

\hypersetup{
    colorlinks=true,
    linkcolor=blue,
    filecolor=blue,      
    urlcolor=blue,
}

\begin{document}

\title{Feedback and ionized gas outflows in four low-radio power AGN at z $\sim$0.15}

\author{L. Ulivi \inst{1,2,3}
\and G. Venturi \inst{5,4,2}
\and G. Cresci \inst{1,2}
\and A. Marconi \inst{1,2}
\and C. Marconcini \inst{1,2}
\and A. Amiri \inst{6}
\and F. Belfiore \inst{2}
\and E. Bertola \inst{2}
\and S. Carniani \inst{5}
\and Q. D'Amato \inst{2}
\and E. Di Teodoro \inst{1,2}
\and M. Ginolfi \inst{2,7}
 \and A. Girdhar \inst{7,8,9}
\and C. Harrison \inst{9}
\and R. Maiolino \inst{10,11,12}
\and F. Mannucci \inst{2}
\and M. Mingozzi \inst{13}
\and M. Perna \inst{14}
\and M. Scialpi \inst{1,2}
\and N. Tomicic \inst{1,2}
\and G. Tozzi \inst{1,2}
\and E. Treister \inst{4}
      %\fnmsep\thanks{Just to show the usageof the elements in the author field}
}
   %1
\institute{Università di Firenze, Dipartimento di Fisica e Astronomia, via G. Sansone 1, 50019 Sesto F.no, Firenze, Italy\\\email{lorenzo.ulivi1@unifi.it}
\and %2
INAF - Osservatorio Astrofisico di Arcetri, Largo E. Fermi 5, I-50125 Firenze, Italy
\and %3
University of Trento, Via Sommarive 14, I-38123 Trento, Italy
\and %4
Instituto de Astrofísica, Facultad de Física, Pontificia Universidad Católica de Chile, Casilla 306, Santiago 22, Chile
\and %5
Scuola Normale Superiore, Piazza dei Cavalieri 7, 56126, Pisa, Italy.
\and %6
Department of Physics, University of Arkansas, 226 Physics Building, 825 West Dickson Street, Fayetteville, AR 72701, USA
\and %7
European Southern Observatory, Karl-Schwarzschild-Str. 2, D-85748 Garching, Germany
\and %8
Ludwig Maximilian Universität, Professor-Huber-Platz 2, 80539 Munich, Germany
\and %9
School of Mathematics, Statistics and Physics, Newcastle University, NE1 7RU, UK
\and %10
Kavli Institute for Cosmology, University of Cambridge, Madingley Road, Cambridge CB3 0HA, UK
\and %11
Cavendish Laboratory, Astrophysics Group, University of Cambridge, 9 JJ Thomson Avenue, Cambridge CB3 0HE, UK
\and %12
Department of Physics \& Astronomy, University College London, Gower Street, London WC1E 6BT, UK
\and %13
Space Telescope Science Institute, 3700 San Martin Drive, Baltimore, MD 21218, USA
\and %14
Centro de Astrobiología, (CAB, CSIC–INTA), Departamento de Astrofísica, Cra. de Ajalvir Km. 4, 28850 – Torrejón de Ardoz,
Madrid, Spain
}
    \titlerunning{Feedback and outflows in jetted AGN}
    \authorrunning{Ulivi, L., et al.}

   \date{Received September 15, 2023; accepted March 16, 2023}

% \abstract{}{}{}{}{} 
% 5 {} token are mandatory
 
  \abstract
  % context heading (optional)
  % {} leave it empty if necessary  
   {An increasing number of observations and simulations suggests that low-power (<10$^{44}$ erg s$^{-1}$) jets may be a significant channel of feedback produced by active galactic nuclei (AGN), but little is known about their actual effect on their host galaxies from the observational point of view.
   We targeted four luminous type 2 AGN hosting moderately powerful radio emission ($\sim$10$^{44}$ erg s$^{-1}$), two of which and possibly a third are associated with jets, with optical integral field spectroscopy observations from the Multi Unit Spectroscopic Explorer (MUSE) at the Very Large Telescope (VLT) to analyze the properties of their ionized gas as well as the properties and effects of ionized outflows. We combined these observations with Very Large Array (VLA) and e-MERLIN data to investigate the relations and interactions between the radio jets and host galaxies. We detected ionized outflows as traced by the fast bulk motion of the gas. The outflows extended over kiloparsec scales in the direction of the jet, when present. In the two sources with resolved radio jets, we detected a strong enhancement in the emission-line velocity dispersion (up to 1000 km s$^{-1}$) perpendicular to the direction of the radio jets. We also found a correlation between the mass and the energetics of this high-velocity dispersion gas and the radio power, which supports the idea that the radio emission may cause the enhanced turbulence. This phenomenon, which is now being observed in an increasing number of objects, might represent an important channel for AGN feedback on galaxies. \\
    
  % aims heading (mandatory)
   
  % results heading (mandatory)
   }
  % conclusions heading (optional), leave it empty if necessary 

 %  \keywords{giant planet formation --
%              $\kappa$-mechanism --
  %              stability of gas spheres
 %              }

    \maketitle
%
%-------------------------------------------------------------------
 \section{Introduction}
Active galactic nuclei (AGN) are thought to play a significant role in shaping the formation and evolution of galaxies, particularly at the high-mass end, through a series of processes known as AGN feedback (\citealt{Silk1998,Fabian2012}). This mechanism is considered an essential element in semianalytic models and hydrodynamic simulations of galaxy evolution because it helps explain a number of the observed properties of galaxies that can otherwise not be accounted for \citep[e.g.][]{KR1995,HR2004, Marconi2004, FM2000,Gebhardt2000,Fabian2012}.\\
AGN feedback is thought to operate in two main modes. In the radiative mode, also known as quasar or wind mode, in which the radiation emitted by the accreting central supermassive black hole (SMBH) drives powerful outflows that can suppress star formation in the host galaxy by removing, displacing, or heating the interstellar medium (ISM; e.g., \citealt{Cicone2014,Harrison2014,Cresci2015a,Carniani2015,DallAgnol}). The second mode is the kinetic mode, also known as radio mode, in which AGN feedback occurs through powerful ($\gtrsim$10$^{45}$ erg/s) extended ($\sim$100s kpc) jets launched directly from the central AGN (\citealt{Fabian2012}). These radio jets are thought to be crucial for AGN feedback because they are able to heat the circumgalactic medium (CGM) and prevent the cooling of gas, thereby suppressing the formation of new stars (e.g., \citealt{McNamara2012}). They are also capable of launching energetic outflows that can remove gas from the host galaxy (e.g., \citealt{Nesvadba2008,Vayner2017}). Recent studies have found from both observational (\citealt{Combes2013,GarciaBurillo2014,Cresci2015a,Jarvis2019,Molyneaux,Venturi2021,Venturi+23, Jarvis2021, Girdhar2022,Audibert2023, Peralta2023}) and theoretical (\citealt{Mukherjee2016,Mukherjee2018,Mukherjee2018a,Mandal2021,Meenakshi2022}) perspectives that low-power (<10$^{44}$ erg s$^{-1}$) and compact jets (lower than 1 kpc) can also have a significant impact on their host galaxy by injecting turbulence in the ISM and accelerating outflows.
\\
\begin{figure*}[t!]
\centering
    \includegraphics[width = 1 \textwidth]{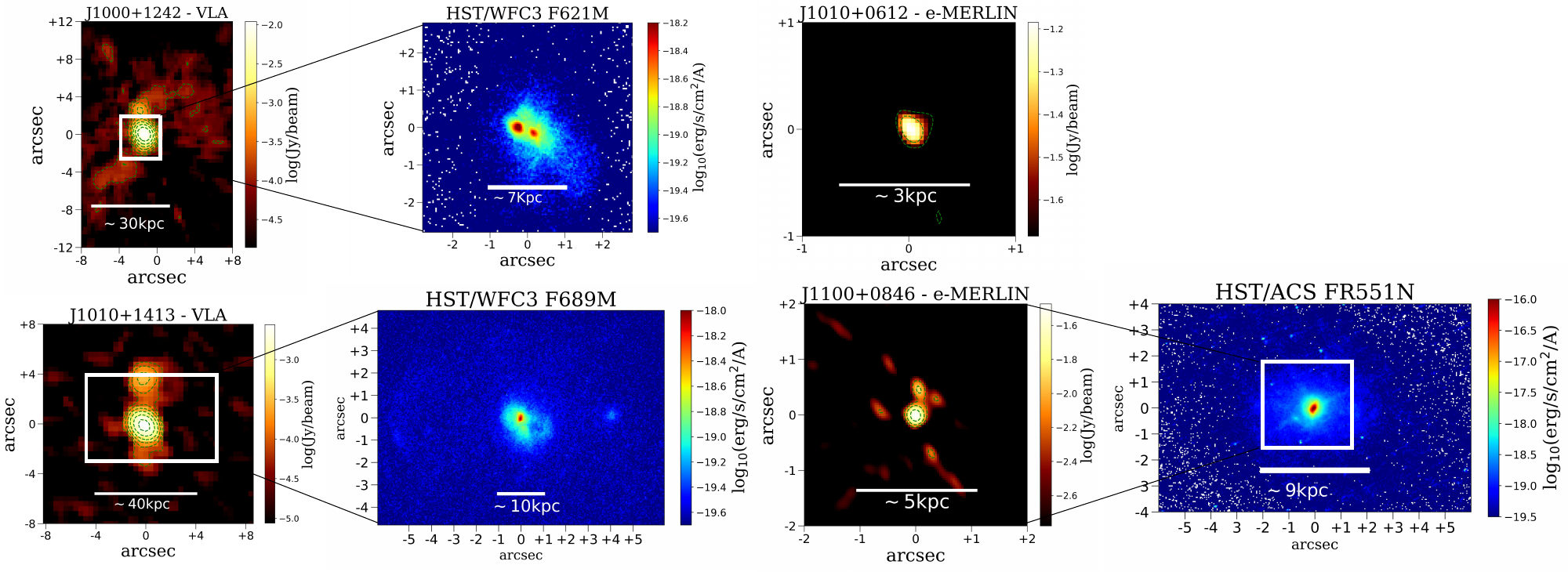}
    \caption{Left panels: Low-resolution ($\sim$1 arcsec HPBW) 6 GHz VLA radio images from \cite{Jarvis2019} for J1000+1242 (top) and J1010+1413 (bottom). Right panels: e-MERLIN data ($\sim$0.2 arcsec HPBW) for J1010+0612 (top) and J1100+0846 (bottom). All panels show a zoomed-in inset with the optical images from \textit{HST}/WFC3. The J1010+0612 radio data are not resolved, and no \textit{HST} data are available for this AGN. The green contours represent [2,4,8,16,32,64,128]$\sigma$ for the VLA images and [8,16,32,64,128]$\sigma$ for e-MERLIN. }  
    \label{HST}
\end{figure*}
From an observational standpoint, classical models of feedback are not always able to fully explain the observational properties of galaxies. This means that it is important to identify and understand new mechanisms of feedback. In a recent study, \cite{Venturi2021} identified a potentially significant mechanism of AGN feedback on the galaxy ISM associated with the presence of low-power (<10$^{44}$ erg s$^{-1}$) compact ($\sim 1$ kpc) radio jets. Local Seyfert galaxies hosting low-power jets oriented at a low inclination relative to the disk of the galaxy have been found to exhibit an increase in the line width of the main emission lines perpendicular to the radio jet, without obvious signs of a strong systematic velocity shift of the line centroid. This enhanced velocity dispersion may be caused by the high levels of turbulence that are injected into the ISM by the jet. Although this phenomenon was previously observed in other local Seyfert galaxies \citep[e.g.][]{Finlez2018,Couto2014,Shimizu2019,Freitas2018,Riffel2015,Bianchin2021,Ruschel-Dutra+21}, the possible connection to the jet was thoroughly investigated, and other explanations were proposed, such as beam smearing, equatorial outflows from the accretion disk, or multidirectional outflows. In some cases, the enhanced perpendicular line width was observed not only in ionized gas, but also in molecular gas, although the enhancement is not as strong as in the ionized phase \citep[e.g.][]{Riffel2015,Shimizu2019,Bianchin2021,Girdhar2022, Audibert2023}. The hypothesis that jet-driven shocks might cause the increase in the velocity dispersion is further supported by recent simulations, which revealed that the interaction between jets and the ISM depends on the properties of the jet, such as inclination and power (\citealt{Mukherjee2018}).
Previous works \citep{Mullaney2013,Molyneaux, Woo2016} have shown that AGN with moderate radio luminosity ($L_{1.4 \rm GHz}$ = 10$^{23-25} ~\rm W ~\rm Hz^{-1}$) can significantly affect the [OIII] profile, leading to an enhancement of the line width. 

The main goal of this paper is to examine the impact of  moderately luminous jets on the host galaxy and to explore the interplay between the AGN and its surroundings. First, we investigate the properties of ionized galactic outflows and the physical conditions of the ionized gas in four type 2 AGN at z $\sim$ 0.15 in detail using spatially resolved optical integral field spectroscopic (IFS) observations from the Multi Unit Spectroscopic Explorer (MUSE) at the Very Large Telescope (VLT). We then compare our MUSE observations with radio-emission maps from the Very Large Array (VLA) and e-MERLIN to investigate the effects of moderate-power radio jets ($\sim 10^{44}$erg s$^{-1}$) on the host galaxy, extending the work of \cite{Venturi2021} on low-power radio (i.e., $\lesssim10^{44}$ erg s$^{-1}$) AGN jets to higher jet luminosities. \\ This paper is structured as follows. We present the sample selection, the data reduction, and the spectroscopic analysis in Sections \ref{Sample} and \ref{Analysis}, respectively. In Section \ref{Results} we present the results of our analysis, including emission-line flux and kinematic maps, gas-excitation diagrams, and a characterization of the outflowing gas through computing their energetics. We then examine the effects of radio jets on the host galaxy in detail and compute the mass and energy of the gas involved in the jet-ISM interaction. We compare our results to those of \cite{Venturi2021} as well as to other results from the literature in Section \ref{discussion}. Finally, we summarize our conclusions in Section \ref{Conclusions}.

\section{Sample selection and data reduction} 
\label{Sample}

\begin{table*}[t!]\small 
\caption{Basic properties of the four AGN in our sample. (1) Right Ascension. (2) Declination. (3) Estimated redshift from \cite{Harrison2014}. (4) Luminosity distance in Mpc taken from NASA/IPAC Extragalactic Database (\url{https://ned.ipac.caltech.edu/}). (5) Starting date of the observations. (6) Mean DIMM seeing measured during the observations at 500 nm and at zenith. (7) Number of exposures. (8) Exposure time per exposure. (9) Total exposure time on target.} \label{sample} \centering 
\begin{tabular}{cccccccccc}
%\begin{tabular}{%
%|>{}p{3cm}%
%|>{}p{2.5cm}%
%|>{}p{25mm}
%|>{}p{15mm}%
%|>{}p{1.5cm}%
%|>{}p{20mm}
%|}
\hline
\hline
Name & RA$^{(1)}$ & Dec$^{(2)}$ & $z^{(3)}$ & $D_{L}^{(4)}$ & Date$^{(5)}$ & DIMM$^{(6)}$ & N frames$^{(7)}$ & $t_\mathrm{exp}^{(8)}$ & $t_\mathrm{tot}^{(9)}$ \\
& [hh:mm:ss] & [dd:mm:ss] & & [Mpc] & [DD-MM-YYYY] & [arcsec] & & [s] & [s] \\
\hline
&&&&\\
J1000+1242 & 10:00:13.14 & +12:42:26.2 & 0.1480 & 732 & 18-02-2020 & 0.73 & 11  & 700 & 7700 \\
&&&&\\
J1010+1413 & 10:10:22.95 & +14:13:00.9 & 0.1992 & 1010 & 29-01-2020 & 0.62 & 8  & 700 & 5600 \\
&&&&\\
J1010+0612 & 10:10:43.36 & +06:12:01.4 & 0.0984 & 468 & 02-04-2020 & 0.78 & 8 & 700 & 5600 \\
&&&&\\
J1100+0846 & 11:00:12.38 & +08:46:16.3 & 0.1005 & 483 & 03-03-2020 & 1.03 & 6 & 450 & 2700 \\
&&&&\\
\hline
\hline
\end{tabular}
\end{table*}
We analyzed MUSE data of four type 2 AGN (J1000+1242, J1010+1413, J1010+0612, J1100+0846; see Table \ref{sample}). These AGN were selected from \cite{Jarvis2019} for their high bolometric luminosity (>10$^{45}$ erg/s; placing these sources in the quasar regime), the extended ionized outflows in [OIII]5007 and the radio emission, with the aim to study the properties of the outflows and the relation between the ionized gas and radio jets, when resolved . The AGN in our sample are classified as radio quiet according to the criterion established by \cite{Xu1999}, which is based on the correlation between the radio luminosity at 5 GHz and the [OIII]5007 luminosity. Despite this classification, they exhibit moderate radio luminosity  ($\sim 10^{24}$ W Hz$^{-1}$ at 1.4 GHz) and host jets in most cases. To study the morphology of the radio emission and its detailed structures, we used the VLA and e-MERLIN interferometric radio observations presented in \cite{Jarvis2019}. In two out of four galaxies (J1000+1242 and J1010+1413), the radio emission that is associated with jets (\citealt{Jarvis2019}) is extended on 10-20 kpc scales. This allowed us to study the relation between the radio jet and the ionized gas in detail. For J1100+0864, only an ambiguous extended feature ($\sim 0.8$ kpc north of the nucleus) is detected with e-MERLIN, while for J1010+0612, the radio emission is not resolved ($\lesssim$ 0.3 kpc).
Consistently with \cite{Jarvis2019}, we computed the noise using 8$\sigma$ clipping repeated ten times in a region 50 times the size of the beam (1'' for VLA and 0.25'' for e-MERLIN.)
For J1000+1242, J1010+1413, and J1100+0846, we also exploited \textit{Hubble Space Telescope} (\textit{HST}) WFC3 and ACS images obtained with the F621M, F689M, and FR551N filters. These filters have a spectral coverage of $\sim$800, 800, and 200 \AA\ around the central wavelengths 6218, 6876, and 5510 \AA, respectively (\citealt{Rodrigo2015}). For the redshifts of our targets, these filters trace the continuum emission of the galaxies. The \textit{HST} images were retrieved from the Hubble Legacy Archive. The radio maps and \textit{HST} images are shown in Fig. \ref{HST}.

\subsection{Data reduction}

The MUSE data of the four galaxies presented in this work belong to ESO program 0104.B-0476 (PI G. Venturi). All the observations were acquired in seeing-limited wide-field mode (WFM), which covers a field of view (FOV) of $1'\times1'$ with a sampling of 0.2 arcsec/spaxel. The data consist of two observing blocks (OBs) for J1010+1413 and J1010+0612 for a total of eight exposures of 700s each, one OB for J1100+0846 for a total of six exposures of 450s each, and five OBs for J1000+1242 for a total of 16 exposure of 700s each; 5 of these 16 exposures were discarded during the data reduction due to the significantly worse seeing (>1''), and we thus combined only 11 exposures to produce the data cube employed in this work for J1000+1242. We estimated the atmospheric optical seeing at the zenith at 500 nm (DIMM) by averaging the mean of DIMM seeing at the start and end of the observation over all the exposures. The main technical details of the observations are reported in Table \ref{sample}. Subsequent exposures were dithered and rotated by 90 degrees in order to remove the artifacts produced by the 24 channels associated with each IFU, as well as to optimize cosmic-ray removal and background subtraction. The sky emission was subtracted using a sky spectrum extracted from regions free of target emission in the science frame.\\ The data reduction and exposure combination were carried out with the ESO MUSE pipeline version v1.6 using ESO Reflex, which provides a graphical and automated way to execute the reduction recipes with ESORex (for details, see \citealt{Weilbacher2020}).

\section{Spectroscopic analysis}
\label{Analysis}

\begin{figure}
    \centering \includegraphics[width = 0.5 \textwidth]{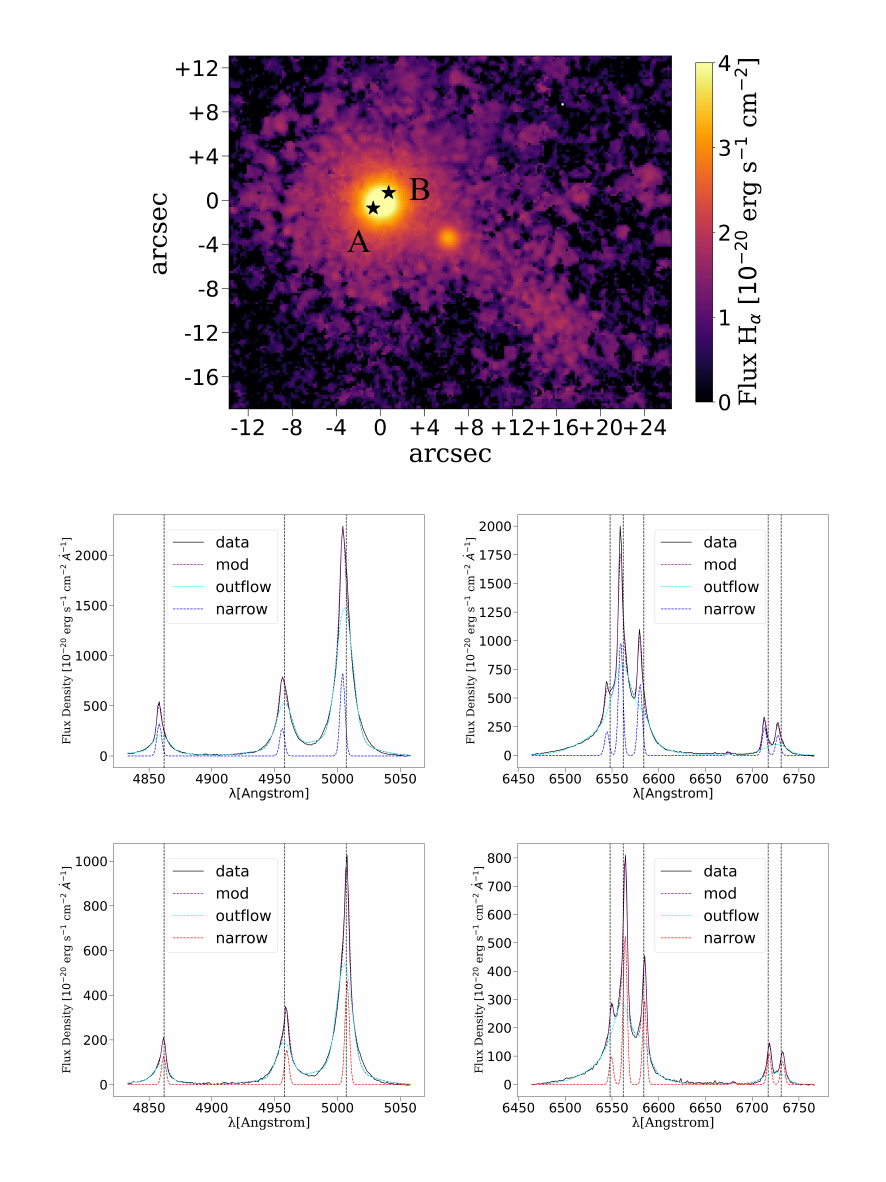}
    \caption{Upper panel: Flux density map of J1010+0612 from MUSE. Bottom panels: Spectra extracted from regions A and B marked with the black stars in the upper panel. The first column shows the spectral range that includes H$\beta$ and [OIII] doublet, and the second column displays the range that includes the H$\alpha$, [NII], and [SII] doublets. The black lines represent the data subtracted from the continuum emission, as described in Section 2,  the purple lines show the total model, the cyan lines show the broad components, and the red (up) and blue (bottom) lines show the narrow components obtained by decoupling the line profiles, as described in Section \ref{Analysis}.}
    \label{Decopuling}
\end{figure}
In this section, we briefly describe the key steps of the analysis of the MUSE optical IFS data cubes to study the emission lines arising from the ISM. The details of the spectral analysis performed in this study can be found in previous works \citep[e.g.][]{Marasco2020,Tozzi2021,Cresci+23}. Before the emission lines were fit, some pre-processing steps were required. First, it was necessary to remove the contribution of the spatially unresolved AGN and the stellar continuum. To do this, we performed a Voronoi tessellation (\citealt{Cappellari2003}) to obtain spatial bins with a signal-to-noise ratio (hereafter, S/N) $>$5 in each wavelength channel (1.25 \AA wide). To remove the continuum, we considered a model that included the continuum emission from the stars and AGN and the line emission modeled with a combination of Gaussian functions. Because our objects are type 2 AGN, we did not include the unresolved emission of the broad line region (BLR) from the core of the AGN in the model. We used the MILES stellar population synthesis templates to model the stellar continuum (\citealt{MILES1,MILES2}). We also adopted a third-order multiplicative polynomial and an additive polynomial of degree between 5 and 7, depending on each target, to reproduce the low contribution of the AGN continuum and the spectral shape, which might be distorted by the reddening. %Regarding the contribution of the emission lines we built a Gaussian model for H$\alpha$, H$\beta$, the oxygen doublets [OIII]$\lambda\lambda4959,5007$ and [OI]$\lambda\lambda6300,6364$, [NII]$\lambda\lambda6548,84$ and [SII]$\lambda\lambda6717,6731 $ with one and with two gaussian components for each line.
We fit the model to the observed spectra employing the penalized pixel-fitting code (PPX; \citealt{Cappellarippxf}). After constructing the total model, we finally subtracted the continuum from the cube spaxel by spaxel and we rescaled the modeled continuum emission obtained in each Voronoi bin to the median of the observed continuum in each spaxel. In this way, we obtained a cube that only contained the contribution of the ionized gas emission lines.
We smoothed each slice of the cube with a Gaussian with a dispersion $\sigma_\mathrm{smooth}$ = 1 spaxel, corresponding to 0.2 arcsec. This allowed us to better investigate the ionized gas by enhancing the S/N while minimizing the degradation of spatial resolution. Then we modeled the emission lines in each spaxel, including H$\alpha$, H$\beta$, the oxygen doublets [OIII]$\lambda4959,5007$ and [OI]$\lambda6300,6364$, [NII]$\lambda6548,6584$, and [SII]$\lambda6717,6731$. We parameterized each line with up to three Gaussian components. To reduce the parameter degeneracies, we kept the velocity and the velocity dispersion of each Gaussian component fixed for all lines, while the flux parameters were left free to vary, with the exception of the [OIII] and [NII] doublets, whose flux ratio was kept fixed to the values imposed by atomic physics (0.343 and 0.338, respectively\footnote{ We retrieved the ratios from \href{NIST}{https://www.nist.gov/pml/atomic-spectra-database} (\citealt{NIST})}). We performed three different fittings by using one, two, and three Gaussian components per emission line. A Kolmogorov-Smirnov (K-S) test on the residuals of the fit was carried out to define the minimum number of components required to provide an acceptable fit in each spaxel (\citealt{Marasco2020}). 
As a final step, we attempted to decouple the kinematics of the ISM by separating the different contributions to the emission line profile (e.g., the rotation or debris merging from the outflows and the gas with high turbulence) 
in order to assign a physical meaning to each component and study each of them separately. 
There is no unique criterion for labeling a single component due to the different nature of the components, but we classified them into narrow or broad components depending on the velocity dispersion of each Gaussian component. Note that ‘broad’ should not be confused with the broad emission lines typical of type 1 AGN, which are emitted from the BLR. For the sources for which  a rotating disk is clearly observed (J1010+0612 and J1100+0846), we set the threshold value for the velocity dispersion above which we classified the components as broad and below which we classified them as narrow, to be 50 km/s higher than the maximum value found for the stellar velocity dispersion (150-200 km/s) to take into account that the ionized gas in the disk could be more turbulent than the stellar kinematics. 
We stress that when fitting three Gaussian components to the emission lines, this approach could result in defining either one component as narrow and two as broad or two as narrow and one as broad.
Fig. \ref{Decopuling} shows a clear example of the spectral decoupling of the emission line profiles between narrow and broad components in J1010+0612. The narrow component represents the emission from the disk while the blueshifted broad component represents the approaching outflow. For J1010+1413 and J1000+1242 we set the thresholds to different values to decouple the line profile, 350 and 250 km/s, respectively, because the stellar kinematics does not show clear rotational features and the ISM kinematics appears to be more complex.

\section{Results}
\label{Results}

\begin{figure*}[t!]
    \centering
    \includegraphics[width = 1 \textwidth]{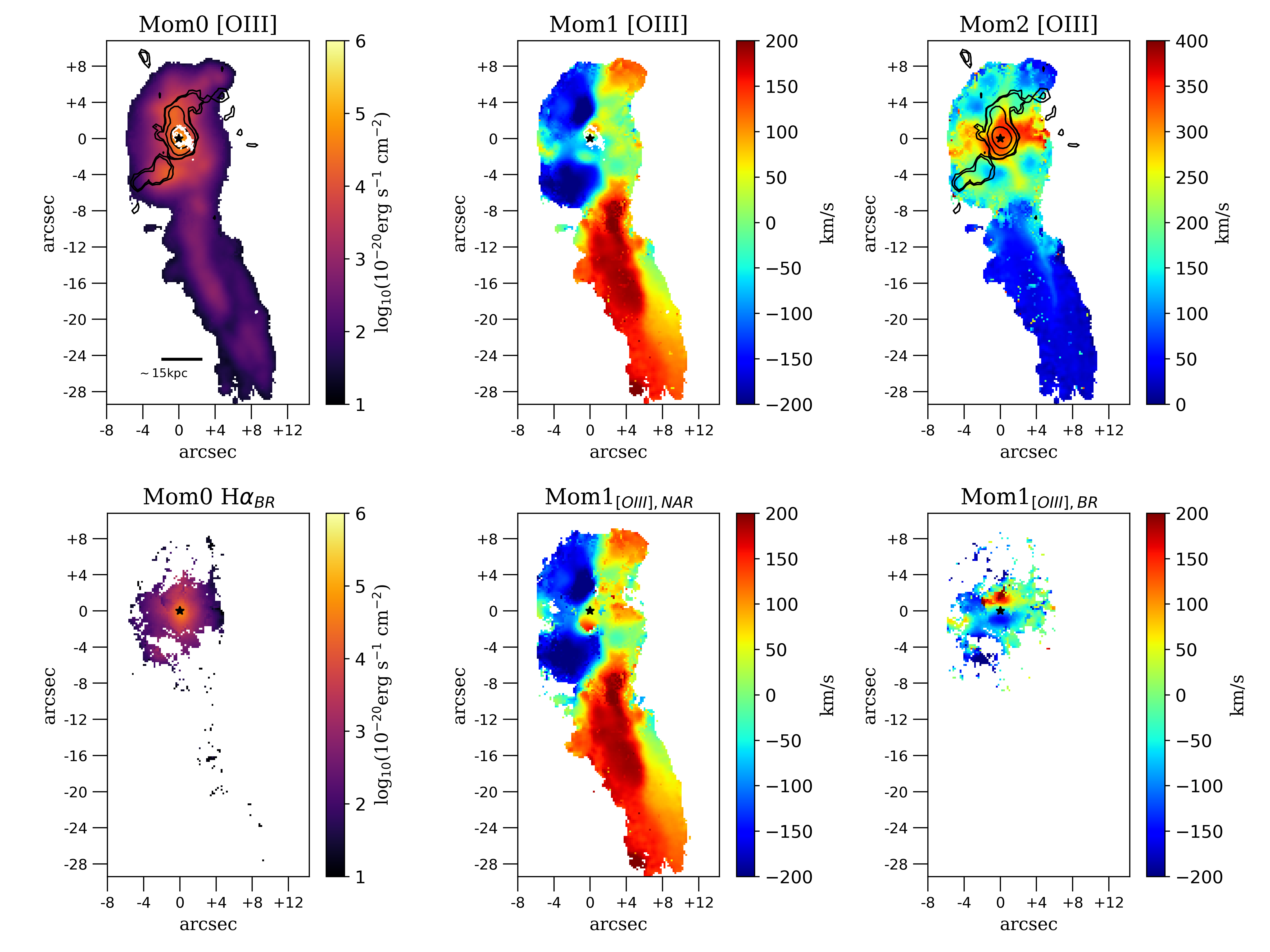}
    \caption{Kinematics of the [OIII] line emission in J1000+1242. The upper panels show the flux (moment 0), velocity (moment 1), and velocity dispersion (moment 2) of the whole line profile. The lower panels show the flux of the broad component of H$\alpha$, the velocity of the [OIII] narrow component (middle), and the velocity of the [OIII] broad component (right) obtained by decoupling the line profile. The black stars indicate the centers of the AGN. The black contours of VLA 6 GHz low resolution ($\sim$ 1'' beam) vary from 0.07 to 2 mJy/beam. The white contours represent the HST continuum. An S/N cut of 3 has been applied to the maps for each line.  }
    \label{MOM012_J1000-1242}
\end{figure*}

\begin{figure*}[t!]
    \centering
    \includegraphics[width = 1 \textwidth]{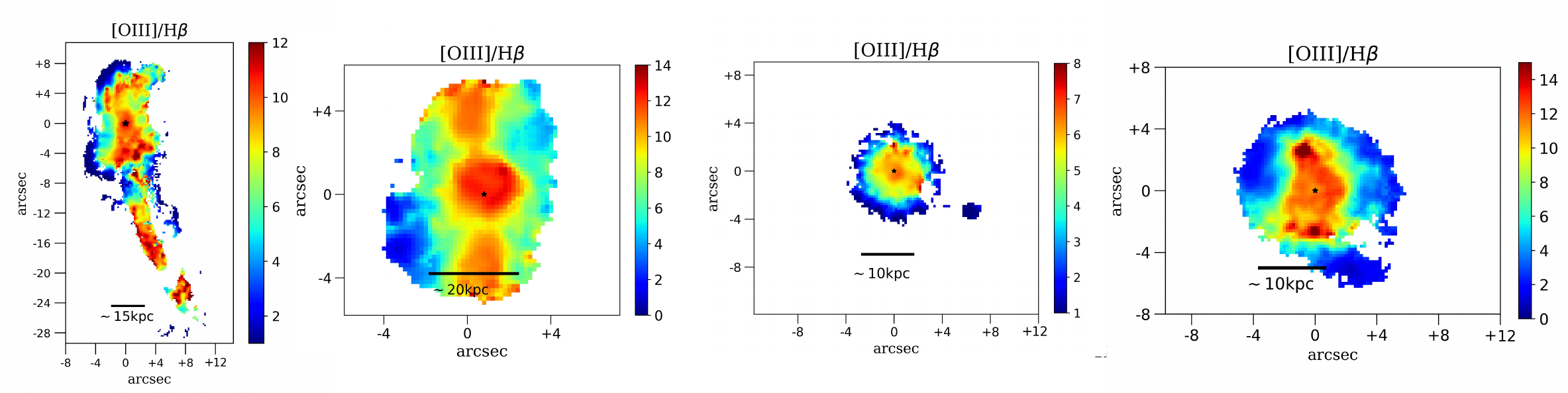}
    \caption{[OIII]/H$\beta$. From left to right: J1000+1242, J1010+1413, J1010+0612, and J1100+0846. An S/N cut of 3 on the flux of each line has been applied to the maps. }
    \label{OIII/HB}
\end{figure*}

\begin{figure}[h!]
    \centering
    \includegraphics[width = 1 \linewidth]{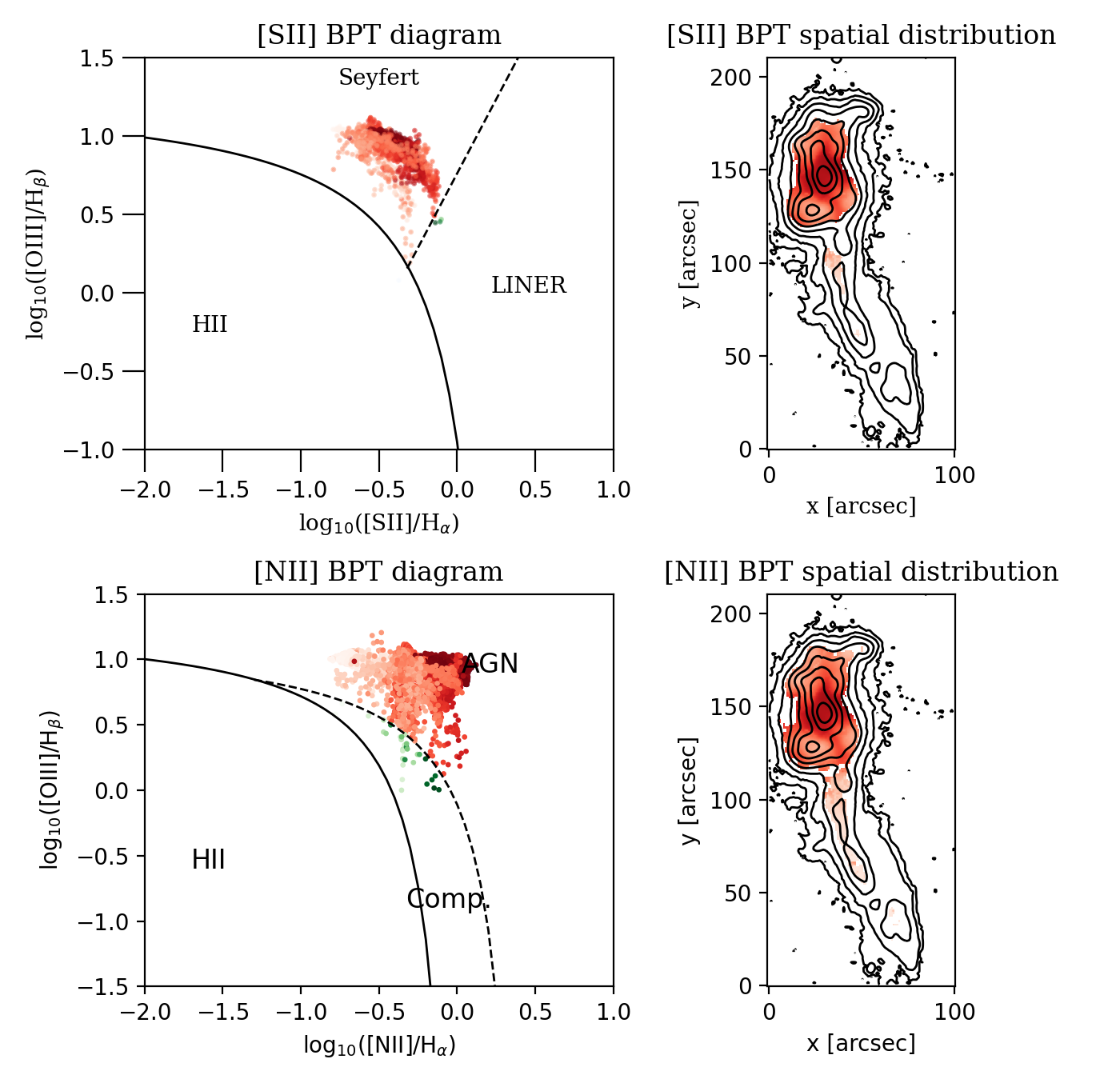}
    \caption{Resolved S-BPT ([OIII]/H$\beta$ vs. [SII]$\lambda$6716,6731/H$\alpha$) (top) and N-BPT ([OIII]/H$\beta$ vs. [NII]$\lambda$6584/H$\alpha$) diagrams (bottom) of J1000+1242 for each spaxel with S/N>3 in each line. The solid curves define the theoretical upper bound for the HII regions (\citealt{Kewley2001}). The dashed line in S-BPT represents the \cite{Kewley} demarcation between Seyfert galaxies and shocks or LINERs.
    The dashed curve in N-BPT represents the empirical classification by \cite{Kauffmann2003} to divide pure star-forming from Seyfert–H II composite objects.
    The black contours represent the moment 0 of H$\alpha$. The data points on the BPT diagram are colored differently to distinguish different regions. The intensity of the color is proportional to the moment 2 of [OIII].}
    \label{BPT_J1000+1242}
\end{figure}

In this section, we present the flux, kinematic, and  emission-line ratio  maps of the ionized gas for the four targets in our sample. 
We mainly focus on the brightest emission line [OIII]$\lambda$5007. We present both [OIII] and H$\alpha$ maps when they show very different characteristics, as in J1100+0846 and J1010+0612. We also present for each target the flux of the broad component of H$\alpha$ that we used to determine the properties of outflowing gas. Moreover, we also provide the moment 0, moment 1, and moment 2 maps, which were calculated on the whole modeled line profile (consisting of either one, two, or three Gaussians). The flux maps were not corrected for the extinction.

\subsection{J1000+1242}
\begin{figure*}[t!]
    \centering
    \includegraphics[width = 1 \textwidth]{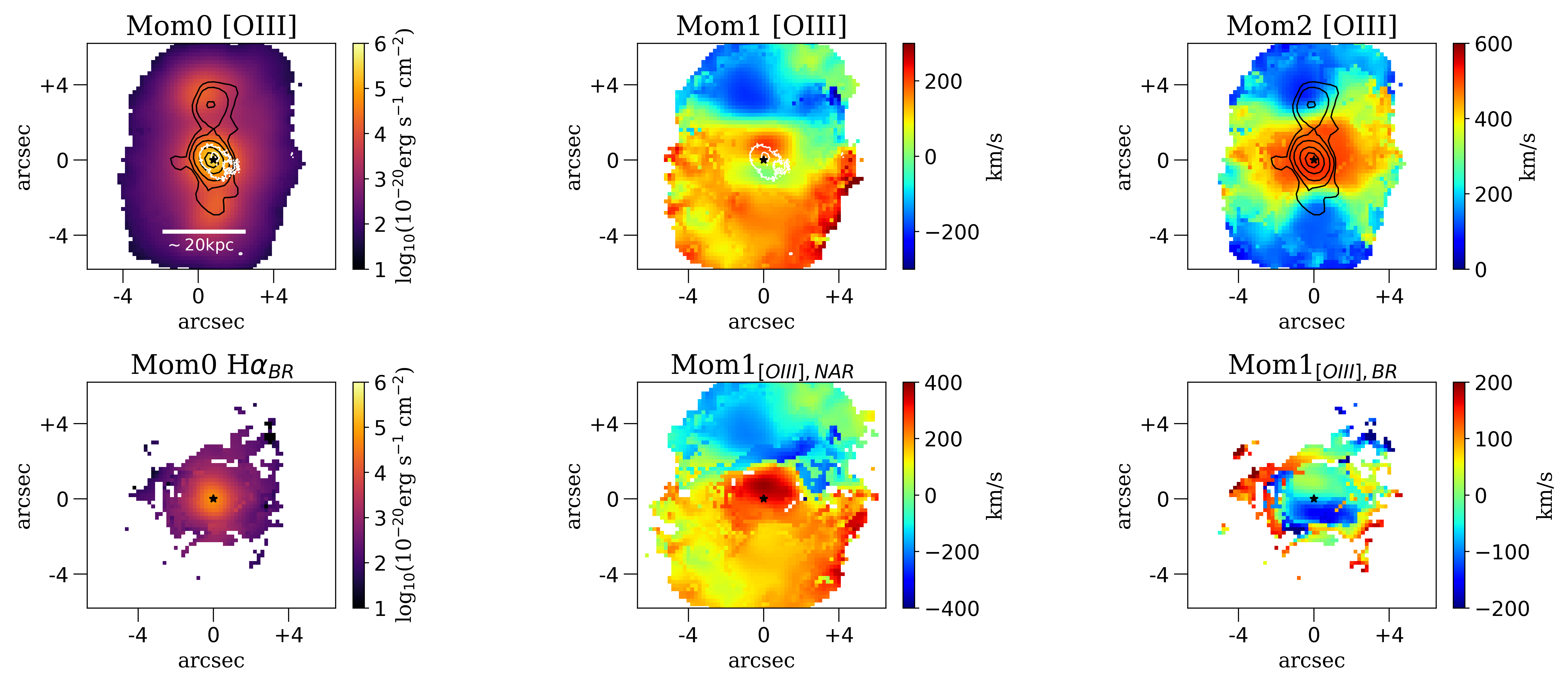}
    \caption{Same as Fig. \ref{MOM012_J1000-1242} for J1010+1413.}
    \label{MOM012_J1010-1413}
\end{figure*}

J1000+1242 is an obscured quasar at z $\sim$ 0.148 (1$''$ $\sim$ 3.5 kpc). The ionized gas morphology shows unambiguous evidence of a merger event, as indicated by the elongated tidal tails (Fig. \ref{MOM012_J1000-1242}). In addition, data from \textit{HST}/WC3 F621M (Fig. \ref{HST}) reveal two distinct nuclear emission sources separated by approximately 1.5 kpc (\citealt{Jarvis2019}). These sources could either represent the double AGN resulting from the merger or the two opposite sides of the NLR obscured by a dusty circumnuclear ring. However, they are not spatially resolved in our MUSE data. An [OIII] bright region tracing the ionized gas is located in correspondence to the radio structures, as found in \cite{Jarvis2019}, suggesting a possible connection between the ionized gas and the radio-emitting regions (as in, e.g., \citealt{Riffel2014},  \citealt{RodriguezArdila+17}, \citealt{May2017,May2018,May2020}, \citealt{Cresci+23}, \citealt{Venturi2021}). The moment 0 map also reveals a tidal tail extending for about 20 arcsec in the southern region. It is predominantly observed in the [OIII] line and might be due to the superposition of two distinct tails. These tails also display distinct kinematic properties. Specifically, the easternmost tail exhibits a velocity of 200 km/s, but in contrast, the westernmost fainter tail shows a velocity from approximately 100-150 km/s in its southern end that gradually decreases towards 0 km/s as it extends northward (Fig. \ref{MOM012_J1000-1242}).
The velocity field of the ionized gas also reveals two distinct blueshifted regions with velocities of $\sim$ -200 km/s located NE and SE of the nucleus, and a redshifted region located NW of it. This velocity field might be due to the twisting of the elongated tidal tails. \\ 
The moment 2 map reveals an increase in the [OIII] velocity dispersion in the region near the nucleus, characterized by an elongated shape oriented W-E and perpendicular to the radio jet (see the radio contours in Fig. \ref{MOM012_J1000-1242}), similar to what was observed by \cite{Venturi2021} and references therein (see Sect. \ref{discussion}). \\ We computed the moment 1 of each spectrally decoupled component (narrow and broad; see Sect. \ref{Analysis}), which are displayed in the bottom panels of Fig. \ref{MOM012_J1000-1242}. The [OIII] broad component shows a velocity gradient of $\sim$350 km/s that extends north-south for 8-10 kpc. It is oriented like the radio jet and consistent with the slight [OIII] line asymmetry seen in \cite{Jarvis2019}. %We identify the region with the velocity gradient of the broad component and with a direction consistent with the radio jet as the region where the outflow is located. \\ 
We interpret the presence of the gradient of the [OIII] broad component, whose direction is consistent with the radio jet, as an outflow launched by the nuclear radio emission, which might demonstrating a direct connection between the radio jet and the outflows.
%\begin{figure}[h!]
%    \centering
 %   \includegraphics[width = 0.5 \linewidth]{Images/Oiii_hb_j1000.png}
%    \caption{[OIII]/H$\beta$ ratio in logaritmic scale (log10). North-south elongated high ionization region is aligned with the direction of the radio jet (white line). Higher [OIII]/H$\beta$ values are found in the center of AGN.}
%    \label{OIII_HB1000}
%\end{figure}
We also present a map of the [OIII]-to-H$\beta$ flux ratio (Fig. \ref{OIII/HB}), which is sensitive to the ionization state of the emitting gas. The map reveals an extended structure, larger than 20 kpc, with high [OIII]/H$\beta$ (larger than 10), oriented in the same direction as the radio jets, indicating the presence of an AGN ionization cone. The value of [OIII]/H$\beta$ remains high at large distances from the nuclear sources. We do not exclude that shocks due to the jet that are cospatial with the ionization cone may also contribute to the ionization of the gas. To verify this, we used the [OIII]/H$\beta$ versus [SII]/H$\alpha$  and [OIII]/H$\beta$ versus [NII]/H$\alpha$ BPT diagrams (e.g., \citealt{BPT, Veilleux1987, Kewley, Lamareille, Law}; see Fig. 4). The map to the right, color-coded according to the location of each spaxel in the diagram to the left, indicates that the main ionization source are the AGN photons. The intensity of the color is proportional to the moment 2 of the [OIII] line. % To better understand the distribution of ionization within the galaxy, we divide the BPT maps into three different sub-regions. 
We find that the ionization cone exhibits higher values of [OIII]/H$\beta$ and lower values of [SII]/H$\alpha$. Conversely, the tidal tails to the south, farther from the nucleus of the AGN, exhibit lower levels of ionization.
%The region perpendicular to the ionization cones falls closer to the low-ionization nuclear emission-line region (LINER) on the BPT diagram, characterized by higher [SII]/H$\alpha$ and lower [OIII]/H$\beta$, which can be associated to the presence of shocks, suggesting that shocks may provide an additional contribution to the ionization in this region.
In the [NII] BPT diagram, the region with a higher velocity dispersion populates the BPT diagram at a higher [NII]/H$\alpha$ ratio, in accordance with a shock scenario  (\citealt{Mingozzi2019,Mingozzi2023}). 

\subsection{J1010+1413}

J1010+1413 ($z\sim0.1992$) is one of the most luminous quasars at $z\sim$ 0.1–0.2 in [OIII] ($L_\mathrm{[OIII]}\sim 1.69\times10^{43}$ erg/s).
The \textit{HST}/WFC3 F621M observation revealed a promising dual AGN candidate with the presence of two distinct [OIII] emitting point sources with a projected separation of 430 pc (\citealt{Goulding2019}). This cannot be probed with our MUSE observations (which have a resolution FWHM of $\sim$ 2 kpc), as in the case of J1000+1242. The morphology and kinematics of J1010+1413 are really complex, as is shown also in the continuum image from \textit{HST}/WFC3 F689M in Fig.\ref{HST}. This is probably due to a recent merger (\citealt{Goulding2019, Jarvis2021}). [OIII] shows strong emission around the nucleus and two lower-luminous [OIII] (as well as in H$\alpha$, which we do not show) hotspots cospatial with the radio lobes (see Figs. \ref{MOM012_J1010-1413} and \ref{HST}). The source is also surrounded by a diffuse halo that is visible in the continuum (Fig. \ref{HST}).
\begin{figure}[h!]
    \centering
    \includegraphics[width = 1 \linewidth]{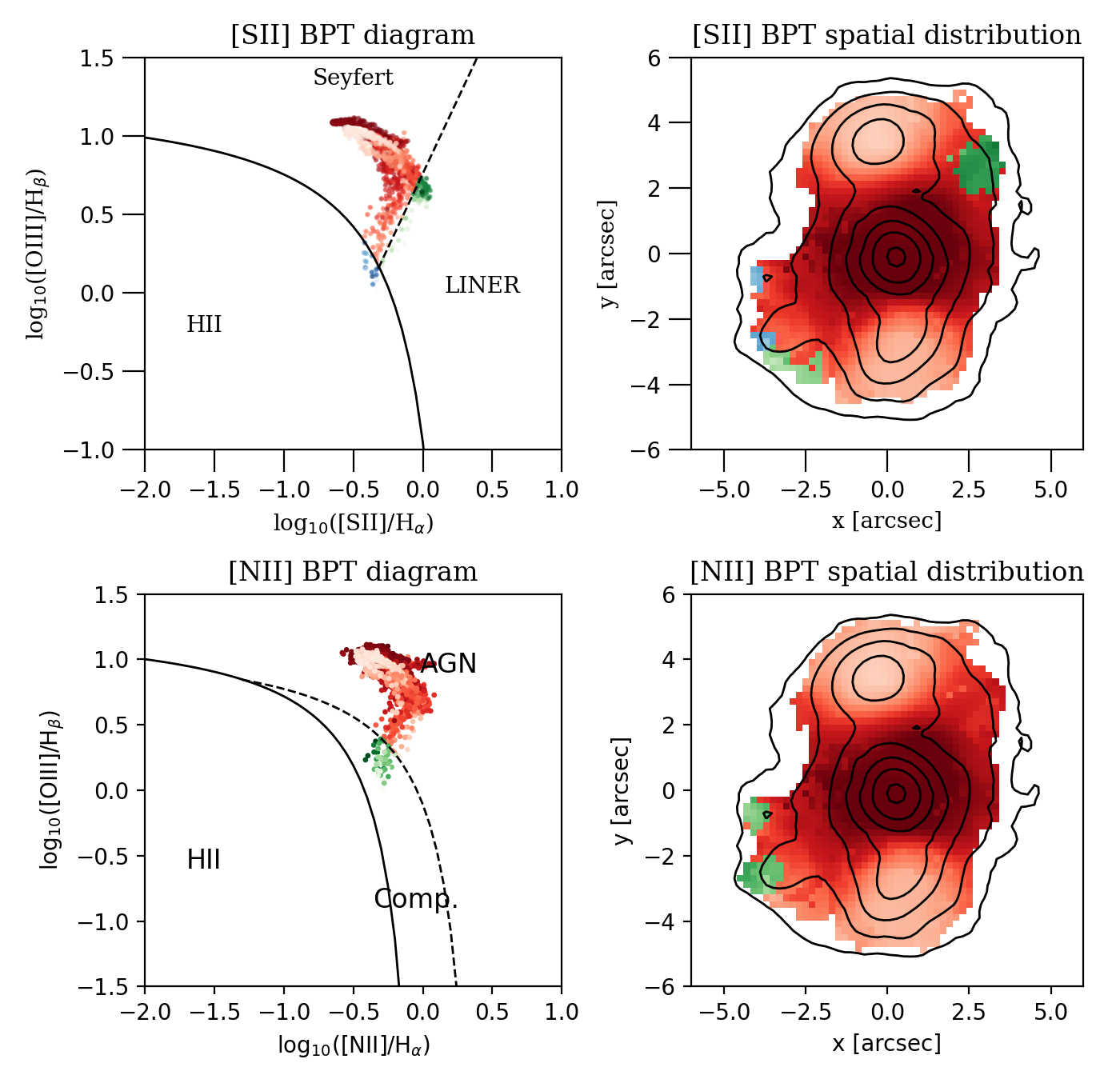}
    \caption{Same as Fig. \ref{BPT_J1000+1242} for J1010+1413.}
    \label{BPT_J1010+1413}
\end{figure}

\begin{figure*}[t!]
    \centering
    \includegraphics[width = 1 \textwidth]{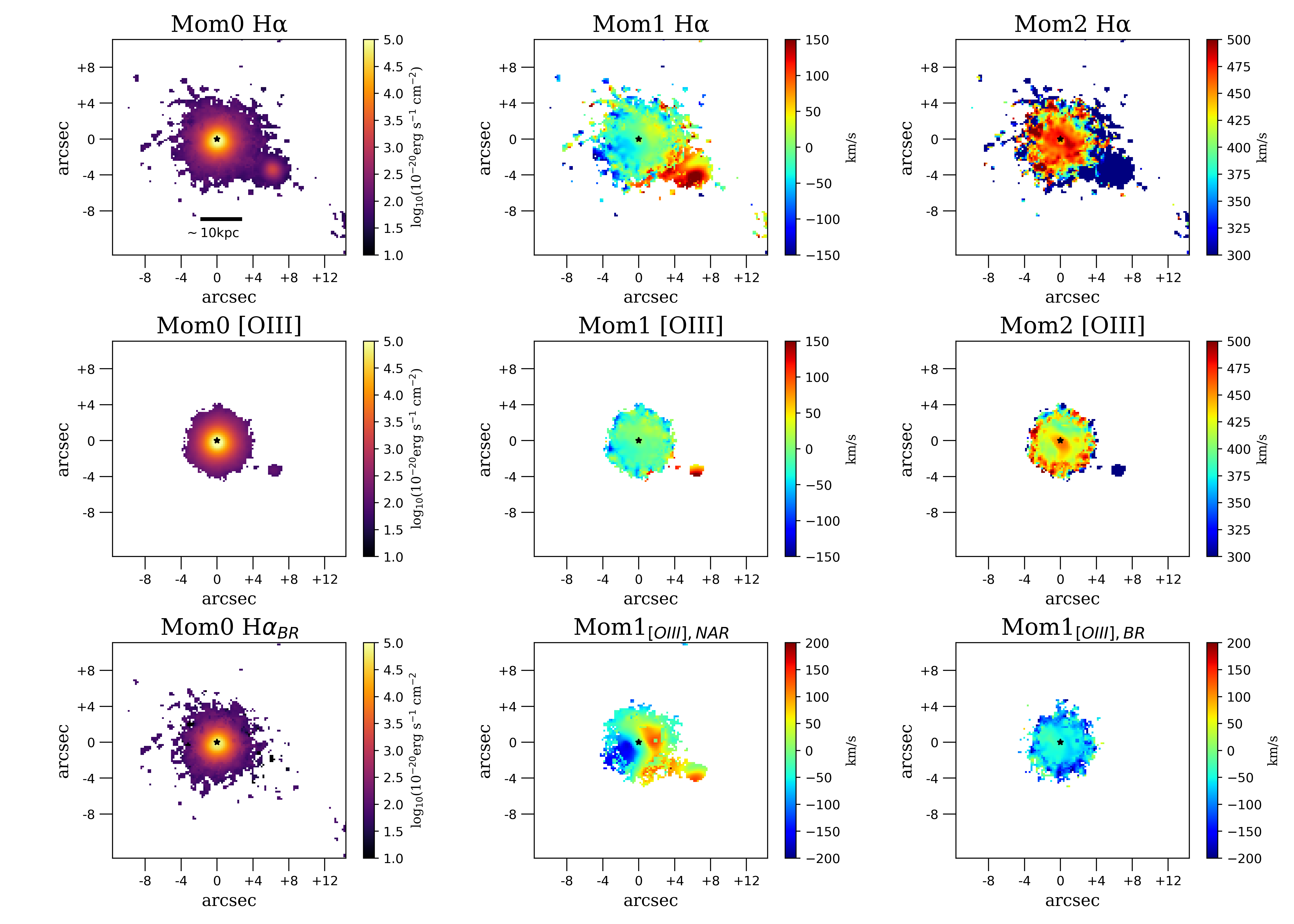}
    \caption{Moment 0, 1, and 2 of Ha (top) and same as Fig. \ref{MOM012_J1000-1242} for J1010+0612.}
    \label{MOM012_J1010-0612}
\end{figure*}
The velocity map of [OIII] reveals two distinct gradients at different scales and with opposite signs. The first gradient is visible at scales of $\sim$2'' around the nucleus in the NS direction, spanning a velocity range of 200 km/s and $\sim$ 0 km/s. The second gradient, at higher scales (5''), corresponds to the northern and southern radio lobes, which are characterized by blue- ($\sim$-200 km/s) and redshifted ($\sim$200 km/s) velocity, respectively.
The [OIII] velocity dispersion map shows that the preferential direction of the high-velocity dispersion ($\sim $ 500 km/s) is aligned perpendicular to the two distinct gradients. This region extends for 20 kpc in the direction perpendicular to the radio jet, as in J1000+1242. Corresponding to the two lobes, the velocity dispersion is lower, as previously observed in other works (\citealt{Shimizu2019, Venturi2021}). The lower velocity dispersion suggests that the lobes could either trace collimated gas that is pushed by the radio jet or rotational gas of the disk that is ionized by the AGN. 
The decoupling of the narrow and broad components was not straightforward in the central region (within 2-3'', i.e., 7 kpc from the nucleus) because all components that contribute to the line profile show a high velocity dispersion, with values over 300 km/s. However, we chose the value of 350 km/s as a reasonable threshold, which allowed us to separate the line profile into a blueshifted broader component (up to 500 km/s), which we identified as the outflow, and a redshifted narrower component (<350 km/s). Because the meaning of this narrower component in this central region is unclear and it has a sigma dispersion of up to 300 km/s at least, we also computed the energetics of the outflow by considering that all the emission is dominated by the outflow, and we also included the regions that correspond to the two radio lobes in the outflow component. 
 %Therefore we applied the \MOKA model presented in Marconcini et al. 2023 where a biconical outflow and a bubble with radial velocity is assumed in order to create a 3D distribution of emitting clouds weighted using the observed line flux and kinematics spaxel by spaxel. For details see Marconcini et al. 2023. 

The maps of gas excitation show a bipolar region with higher [OIII]/H$\beta$ and lower [SII]/H$\alpha$ compared to the regions outside of the bicone (see Fig. \ref{OIII/HB} and Fig. \ref{BPT_J1010+1413}). The regions northwest and southeast of the center have a lower ionization level with a higher value of [SII]/H$\alpha$ and lower value of [OIII]/H$\beta$. Moreover, regions with a lower velocity dispersion populate the BPT diagram at lower [NII]/H$\alpha$ ratio, as was also found in J1000+1242.

%\begin{figure}
 %   \centering
 %   \includegraphics[width = 0.25 \textwidth]{Total_Images/Screenshot 2023-02-27 at 12.27.43.png}
 %   \includegraphics[width = 0.25 \textwidth]{Total_Images/Screenshot 2023-02-27 at 12.28.47.png}
 %   \caption{\MOKA model for J1010+1413} 
 %   \label{fig:my_label}
%\end{figure}

\subsection{J1010+0612}

J1010+0612 is the nearest quasar of our sample with z $\sim$ 0.098(1$''$ $\sim$ 2.3 kpc). The [OIII] line emission is detected in a region limited to the central 7-8 kpc from the nucleus, beyond which the [OIII] flux becomes very faint. The H$\alpha$ flux is more extended and shows a companion galaxy located approximately 16 kpc away (Fig. \ref{MOM012_J1010-0612}, the left panels in first two rows). In contrast to CO observations with ALMA, no double peak is observed in the moment 0 map of H$\alpha$ and [OIII] (\citealt{Ramos22}). The kinematic map of H$\alpha$ (Fig. \ref{MOM012_J1010-0612}, central panel) reveals a blueshifted emission SE (-50 km/s) of the nucleus and a velocity close to $\sim$ 50 km/s NW of the nucleus. This rotational pattern closely resembles that observed in the stellar velocity map (see Appendix A), but with lower absolute values. This discrepancy arises because the line profile, as shown in Fig. \ref{Decopuling}, is not just composed by the disk emission, but also includes a significant contribution from a blue wing, which contributes to blueshifting the global velocity in the region where the stellar component is receding. 
Unlike J1000+1242 and J1010+1413, it is not possible to clearly resolve a precise region with a higher-velocity dispersion, but the motion of the gas is highly disturbed overall throughout the central region (up to 10kpc) of J1010+0612, showing a velocity dispersion of up to 500 km/s. In contrast, the moment 2 of the companion galaxy decreases, suggesting that the merging is still in an early stage.
The two maps displayed in the bottom center and bottom right panels of Fig. \ref{MOM012_J1010-0612} show the velocity maps of the narrow and broad components, respectively. The velocity map of the narrow emission ($V_\mathrm{[OIII],NAR}$) shows a similar rotation as that detected in the stellar kinematics (see Appendix A), providing clear evidence that the narrow emission traces the ionized hydrogen from the disk. The same rotation pattern is also detected in CO(2-1) (\citealt{Ramos22}). The velocity map of the broad emission instead exhibits a blueshifted emission ($V_\mathrm{[OIII],BR}\sim$100 km/s) that we identify as the outflow.  
The resolved S-BPT and N-BPT show that the central region is completely ionized by the AGN, with [OIII]/H$\beta$ up to 8. The companion galaxy is primarily ionized by HII, as shown from S-BPT, and it falls into the composite region in the N-BPT. %Although less clear than the other two galaxies, also in J1010+0612 regions with higher velocity dispersion populate regions with higher [NII]/H$\alpha$.}  
\begin{figure}[h!]
    \centering
    \includegraphics[width = 1 \linewidth]{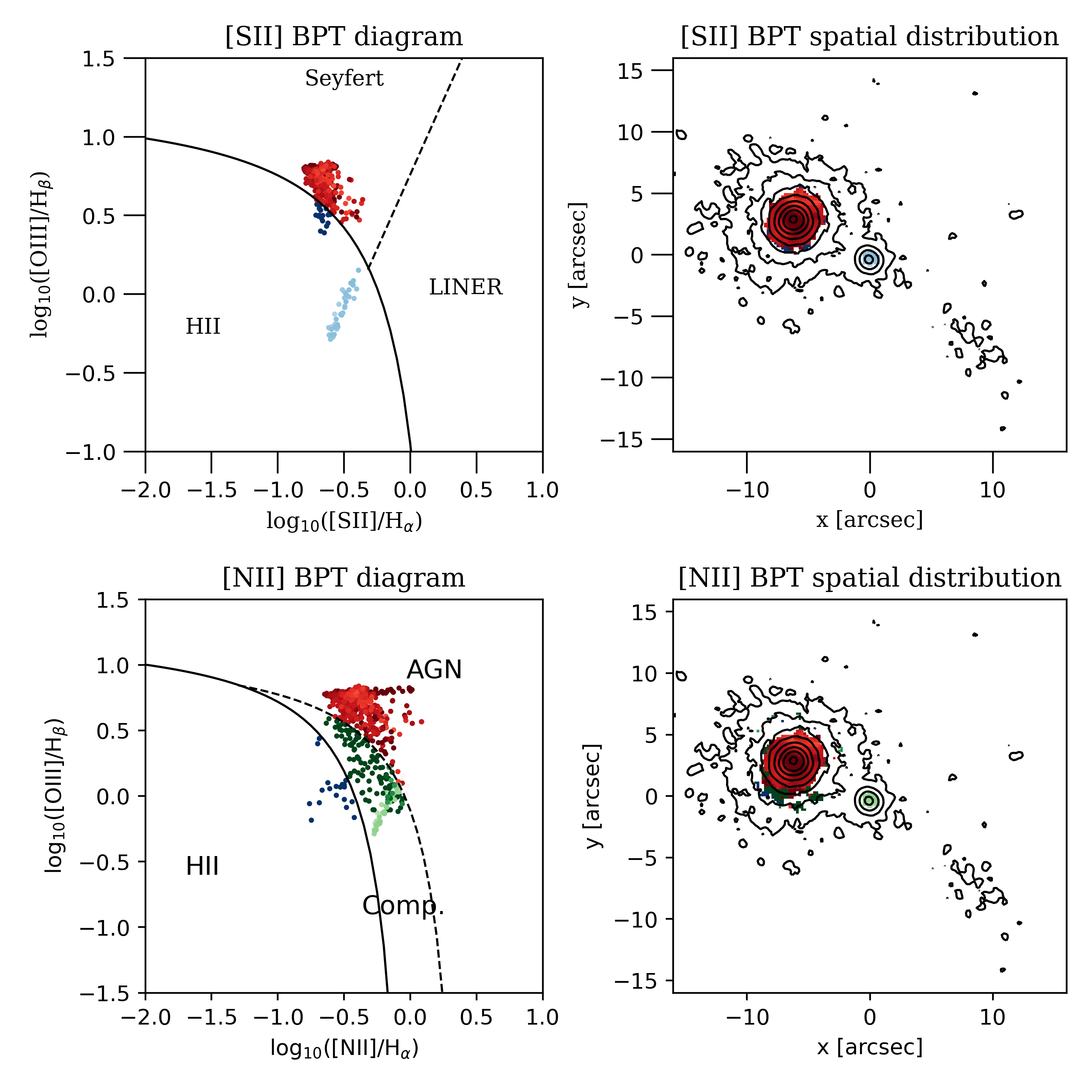}
    \caption{Same as Fig. \ref{BPT_J1000+1242} for J1010+0612}
    \label{BPT_J1010+0612}
\end{figure}

\subsection{J1100+0846}
\begin{figure*}[t!]
\centering
    \includegraphics[width = 1 \textwidth]{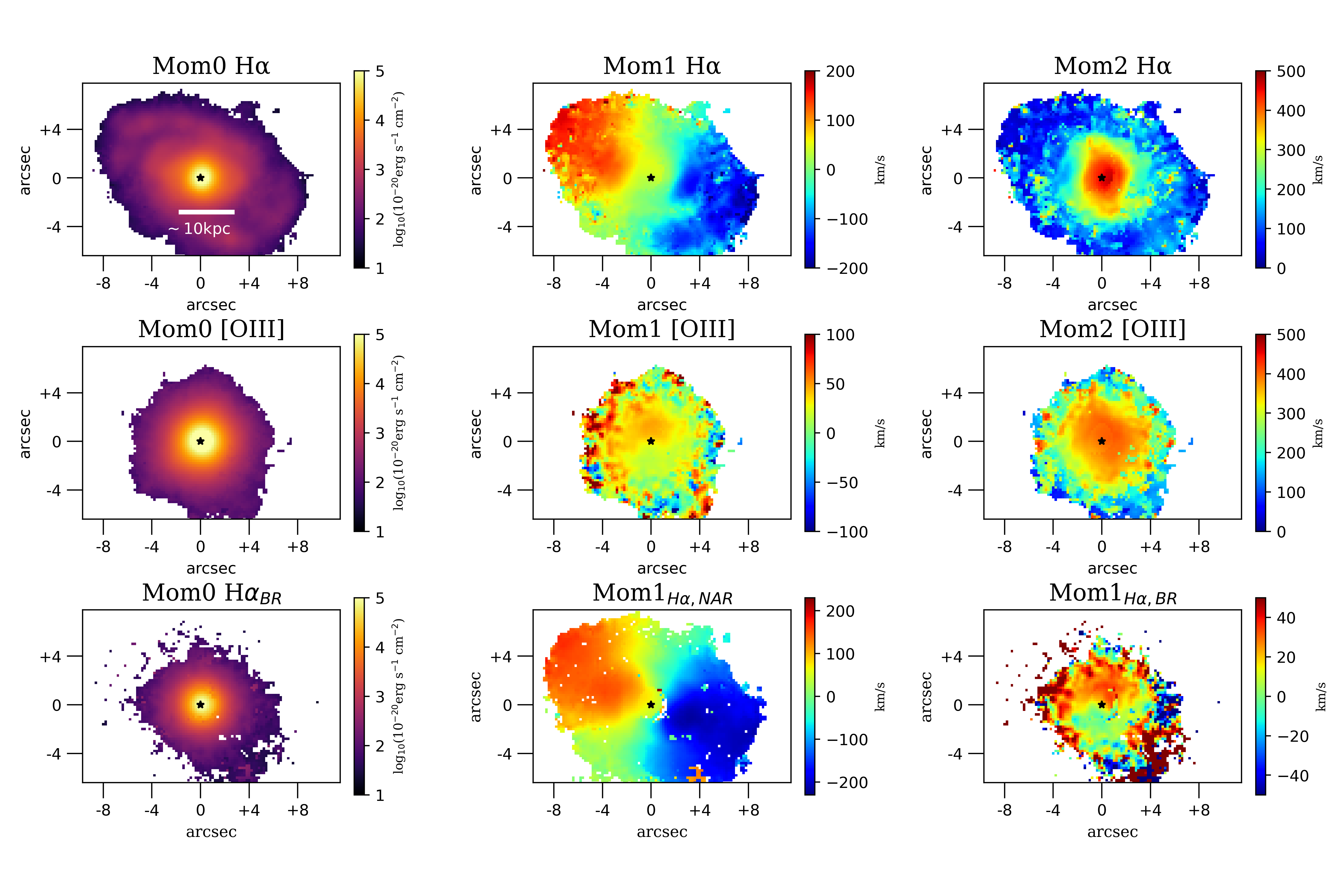}
    \caption{Same as Fig. \ref{MOM012_J1010-0612} for J1100+0846.}
    \label{MOM012_J1100-0846}
\end{figure*}
J1100+0846 is a type 2 AGN that is hosted in a barred spiral galaxy. The high-resolution \textit{HST}/WFC3 image (Fig. \ref{HST}) reveals nuclear elongated emission ($\sim$ 0.4'', i.e. $\sim$ 0.8 kpc) in a direction consistent with the radio structure identified by e-MERLIN.   
Different from the other galaxies, the [OIII] and H$\alpha$ emissions show a different morphology (Fig. \ref{MOM012_J1100-0846}, top left and middle left panels). The H$\alpha$ traces the disk and spiral arms of the galaxy, while the [OIII] emission shows a broad circularly symmetric emission around the nucleus that is only partially resolved. Clear differences in the two emissions are also observed in the moment 1 maps (Fig. \ref{MOM012_J1100-0846}, top center and middle center panels). The integrated H$\alpha$ velocity map mainly traces the rotation of the disk, while the [OIII] velocity is low and close to the systemic velocity of the galaxy. The moment 2 maps reveal that in correspondence with the spiral arms, the H$\alpha$ velocity dispersion is low ($\sim$120 km/s), while in the central region, the gas shows high turbulence with velocity dispersions of up to 500 km/s in both [OIII] and H$\alpha$ (Fig. \ref{MOM012_J1100-0846}, top right and middle right panels). However, it is unclear whether the high-velocity dispersion region lies perpendicular to the direction of the jet (determined by the nucleus and the bright hotspot located N/NW of the nucleus; see Fig. \ref{HST}), as it is in J1000+1242 and in J1010+1413.
We find that the $V_\mathrm{H\alpha,NAR}$ map is considerably smoother than the H$\alpha$ integrated velocity ,and it almost perfectly matches the rotating stellar velocity V$_\mathrm{STAR}$ (see Appendix A). The rotation pattern is consistent with ALMA observations of CO(2-1) in this case as well (\citealt{Ramos22}). $V_\mathrm{H\alpha,BR}$ resembles the first moment of [OIII]. This suggests that this high-velocity dispersion emission dominates the [OIII] line profile. We identify this emission as an outflow with a direction roughly in the plane of sky, given its low velocities in the line of sight (Fig. \ref{MOM012_J1100-0846}, middle center panel).

The S-BPT diagram reveals a constant value of [SII]/H$\alpha$ in each spaxel, and the [OIII]/H$\beta$ ratio spans a wide range of values, showing the highest values in the N-S direction, as also shown in Fig.\ref{OIII/HB}. The ionization is dominated by AGN photons in most of the regions of the target. The same region populates the composite part of the N-BPT diagram.
\begin{figure}[h!]
    \centering
    \includegraphics[width = 1 \linewidth]{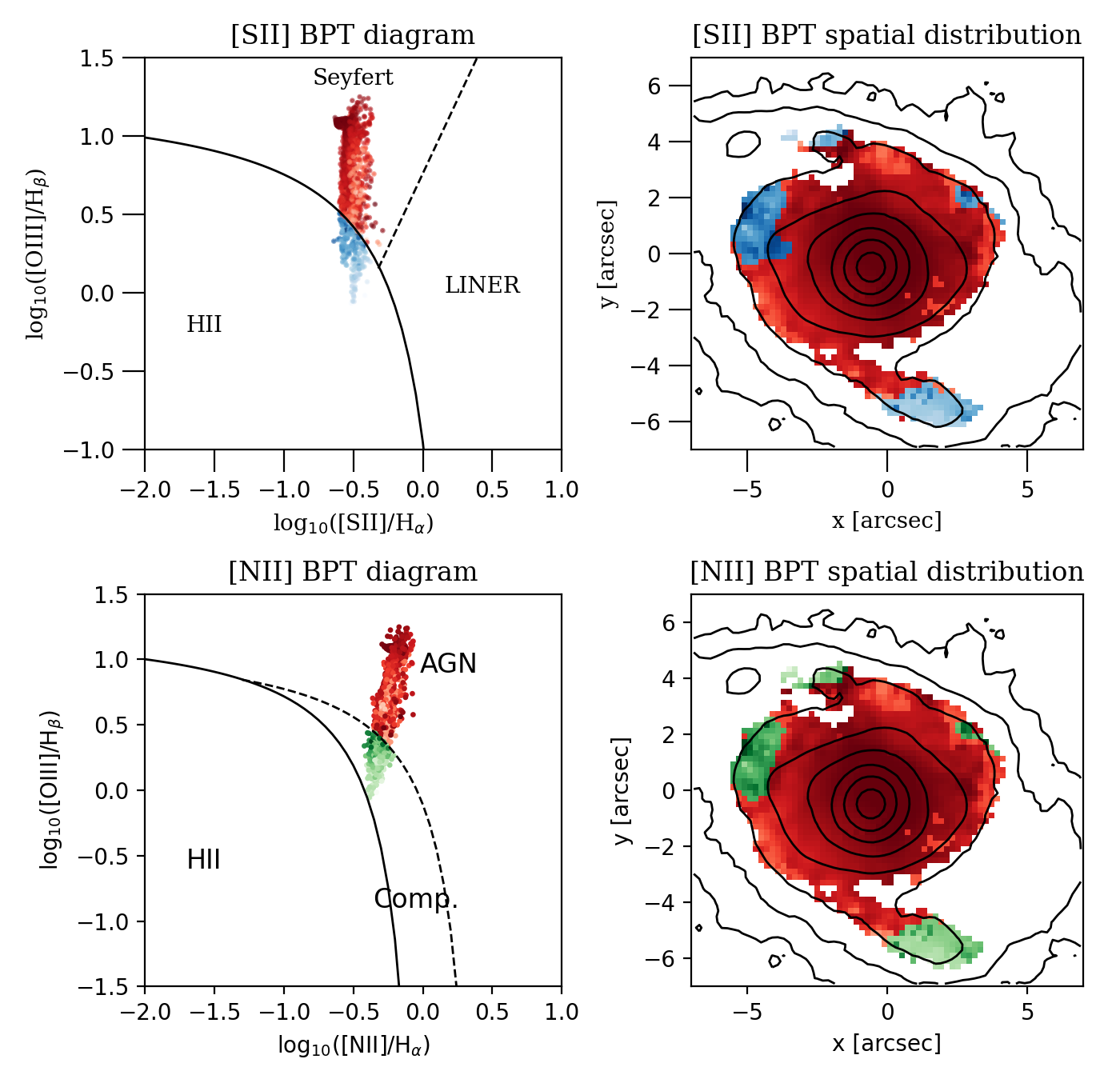}
    \caption{Same as Fig. \ref{BPT_J1000+1242} for J1100+0846.}
    \label{BPT_J1100+0846}
\end{figure}

\subsection{Extinction maps}
\begin{figure*}
    \centering
    \includegraphics[width = 1 \textwidth]{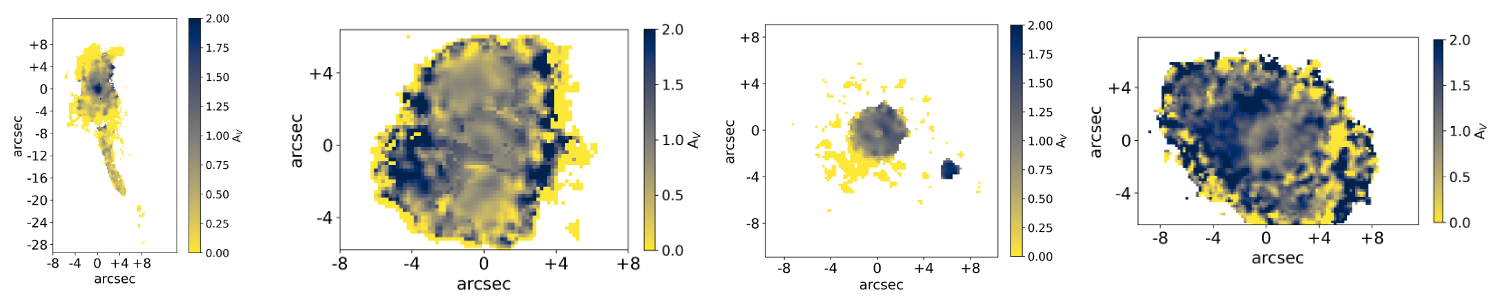}
    \caption{ Extinction maps from the Balmer decrement H$\alpha$/H$\beta$ (in units of magnitude). From left to right: J1000+1242, J1010+1413, J1010+0612, and J1100+0846. An S/N cut of 3 on the flux of each line has been applied to the maps.}
    \label{Extinction}
\end{figure*}

To derive the extinctions, we measured the H$\beta$/H$\alpha$ line ratios from each spaxel and scaled them to calculate $E(B-V)$ as follows (\citealt{Dominguez2013}):
\begin{equation}
E(B-V) = \frac{2.5}{k(\lambda_\mathrm{H\beta})-k(\lambda_\mathrm{H\beta})}\log_{10}\left(\frac{\mathrm{(H\alpha/H\beta)_{obs}}}{\mathrm{(H\alpha/H\beta)_{in}}}\right),
\end{equation}
where k$_{H\beta} \sim 3.7$ and k$_{H\alpha}\sim 2.5$ are the values of the extinction curve by \cite{Cardelli1989} at the H$\beta$ and H$\alpha$ wavelengths, respectively. We assumed $\mathrm{(H\alpha/H\beta)_{in}}$ = 3.1, typical of AGN (\citealt{Veilleux1995}). $A_{V}$ is then calculated as $R_{V}\times E(B-V)$, where $R_{V}$ (3.1) is the selective extinction, and it depends on the physical properties of the dust grains.
We show the extinction maps obtained without separating the kinematic components in Fig. \ref{Extinction}.
 %We found values for A$_{V}$ ranging between 0 and 2. The mean extinction for all targets varies between 1 and 1.8.
The highest value of A$_{V}$ in J1000+1242 is found in the nuclear region ($\sim$2.5 mag ). For J1010+1413 and J1010+0612, we find almost constant values ($\sim$1.5 mag), except for a peak of the extinction in the southeast region of J1010+1413, which we previously identified as a region with lower ionization. For J1010+0846, we obtain a peculiar map of the extinction in which the higher values are found within 2'' and 4'' from the nucleus in the eastern region, corresponding to the higher molecular content of CO(2-1) found in \cite{Ramos22}. The mean extinction for all targets varies between 1 and 1.8 mag. 
We used these maps to correct for extinction the emission line fluxes in our targets spaxel by spaxel. The corrected maps are not reported in this paper, but we used them to compute the properties of the ionized gas.

\subsection{Properties of ionized outflows}
The kinematic structures of the outflows in our objects are varied and complex. To characterize the physical properties of the outflows, we adopted the simplified model by \cite{Genzel2011}, which is valid for case B recombination of fully ionized gas with $T\sim10^{4}$ K.\\
The H$\alpha$ line luminosity can be expressed as
\begin{equation}
    L_\mathrm{H\alpha}=\int_{V}f \  n_{e} \ n_{p} \ j_\mathrm{H\alpha}(n_{e},T_{e}) \ dV
    \label{lum ha},
\end{equation}
where $f$ is the filling factor, which is assumed to be 1, $n_{e}$ is the electron density, $n_{p}=1.2^{-1}\times n_{e} $ is the proton density assuming 10\% number densities of He, and $j_\mathrm{H\alpha}$ is the emissivity of H$\alpha$ %($\sim$3.5$\times10^{-25}T_{4}$erg cm$^{-3}$ s$^{-1}$)
at $T$ = 10$^{4}$ K.\\
The mass of the outflowing gas is given by 
\begin{equation}
     M_\mathrm{out}/M_{\odot} = 3.2 \times 10^{5} \left(\frac{L_\mathrm{H\alpha}}{10^{40}~\mathrm{erg~s^{-1}}}\right)\left(\frac{n_{e}}{100~\mathrm{cm^{-3}}}\right)^{-1}
     \label{Cresci 2017_f},
\end{equation}
where we assumed no dependence between $j_\mathrm{H\alpha}$ and $n_{e}$.

%We computed the mass of the outflow as:
%\begin{equation}
%    M_{out} = 5.33 \times 10^7  \frac{C \ L_{44}([OIII])}{<n_{e3}> \ 10^{[O/H]}} \ M_{\odot} \label{m_out_final_canodiaz} \,
%\end{equation}\\
%where $L_{44}([OIII])$ is the luminosity of the broad component of [OIII] line that we classified as outflow, in units of $10^{44} \ erg/s$, $<n_{e3}> = \int_V n_e \ f \ dV / \int_V  \ f \ dV$ is the average electron density in the ionized outflow gas, $10^{[O/H]}$ represent the oxygen abundance in solar units and $C = <n_{e3}>^2 / <n_{e3}^2>$ is a “condensation factor” that we assumed equal to 1 with the assumption that all ionizing gas clouds have the same density. Also, under these assumptions, the mass of outflowing ionized gas is independent of the filling factor of the emitting clouds.\\
We determined the mass of the ionized outflows for each spaxel by considering in Eq. \ref{Cresci 2017_f} only the contribution of the broad component of H$\alpha$. Then we summed all the contributions from all the spaxels to obtain the total mass of the outflow.  
The total outflow masses $M_\mathrm{out}$ derived for our sample range between $1 \times 10^{7}$ M$_{\odot}$ and $1 \times 10^{8}$ M$_{\odot}$. Assuming a simplified model of a conical outflow with an opening angle $\Omega$, a radially constant mass-loss rate, and an outflow velocity $v_{out}$, out to a radius $R$, we can compute the mass outflow rate as

\begin{equation}
    \dot M_\mathrm{out} = C\frac{M_\mathrm{out}v_\mathrm{out}}{R},
    \label{mout}
\end{equation}
where C depends on the adopted outflow history. In our case, we adopted a constant mass-outflow rate that led to C = 1 (\citealt{Lutz2020}).
As an estimate of the radius of the outflow, we assumed the maximum extension of the blueshifted gas in the broad [OIII] component, defined as the distance at which the line flux reaches 10\% of its total flux. We note that in all of our targets, this radius exceeds the seeing of the observations by more than three times.   %In J1000+1242 and J1010+1413 we considered the radius of the outflow as the distance within the velocity gradient of the broad component is observed.
%Wherever this gradient is not observed (J1010+0846 and J1010+0612) we estimated the radius as the maximum distance from the center where the broad emission ([OIII] or H$\alpha$) reaches at least 10\% of the total intensity of the line.
To estimate the electron density of the outflow, we employed the optical [SII]$\lambda$6717/[SII]$\lambda$6731 ratio using the \cite{OstFer2006} theoretical relation, assuming a typical electron temperature of 10$^{4}$ K. We computed the density spaxel by spaxel using the broad component of [SII] resulting from the decoupling of the line profiles. To enhance the S/N, we integrated the flux of the [SII] doublet over the spaxels for which a physical density could be extracted, and we calculated an average density for the outflow. In the assumed scenario, the average volume density of the outflowing gas is proportional to $R^{-2}$, but in our case, we considered a mean value for the density.  %To minimize the blending, we extracted, if possible, the integrated spectrum of both side of the outflow individually, avoiding the central region where lines are too broad (Nesvadba 2008). 
We used the \textsc{PYNEB}(\citealt{Luridiana2015}) software package to convert the doublet ratio into an electron density. The values of the outflow densities are reported in Table \ref{result_fiore}.
To be consistent with other results in the literature (\citealt{Fiore2017,Fluetsch2019}), we assumed that the bulk wind velocity is a combination of the velocity and the width of the outflow Gaussian component(s). In particular, we adopted the definition of the velocity in \cite{RV2013} as
\begin{equation}
    v_\mathrm{out} = v_\mathrm{broad} + 2\sigma_\mathrm{broad}
    \label{vmax},
\end{equation}
where $v_\mathrm{broad}$ is the central velocity of the broad component compared to the systemic velocity, and $\sigma_\mathrm{broad}$ is the velocity dispersion of the broad component. The concept is that when the line profile of the gas moves away, this appears as the combination of different velocities that are inclined at different angles with respect to our line of sight. Consequently, only the highest observed velocities represent the actual velocity of the outflowing material. We computed it in each spaxel in which the outflow was located, and we then calculated the average of the outflow velocities. We then assigned the standard deviation of the values in that region as the uncertainty on v$_\mathrm{out}$.
%For J1010+0612 and J1100+0846 we also tested the 3D AGN outflow model MOKA 3D presented in Marconcini et al. 2023 where a rotating disc velocity field and a biconical outflow is assumed in order to create a 3D distribution of emitting clouds weighted using the observed line flux and kinematics spaxel by spaxel. For details see Marconcini et al. 2023. \\

We also computed the kinetic energy and the kinetic power of the outflow as
$E_\mathrm{out} = \frac{1}{2} M_\mathrm{out}v_\mathrm{out}^{2} $ and  $    \dot E_\mathrm{out} = \frac{1}{2}\dot M_\mathrm{out}v_\mathrm{out}^{2}$, respectively. Their values are reported in Table \ref{result_fiore} (see also \citealt{Speranza+23} for a comparison of outflow properties in J1010+0612 and J1100+0846). We also estimated the kinetic coupling efficiency, defined as $\epsilon = \dot{E}_\mathrm{out}/L_\mathrm{bol,AGN}$, finding values $\sim10^{-3}-10^{-4}$, in accordance with values found in the literature (\citealt{Costa2018,H2018}).
%A$_{V}^{2}$

%\section{Jet-ISM interaction}

%\subsection{What drives outflows?}
%In two of our targets, J1000+1242 and J1010+1413, we detected a velocity gradient (200-300km/s) of the broad component which we have identified as an outflow. As shown in fig.\ref{sigma_explanation} for J1000+1242, the direction of the outflow is aligned with the jet (purple line) and with the ionization cone (dashed black line). In J1010+0612 and J1100+0846 is ambiguous due to lack of clear and resolved jet features. In J1010+0612 the outflow is blue-shifted and directed toward the observer as well as the unresolved nature of the jet could be attributed to its inclination and be aligned with the outflow. This has led us to wonder whether the jet could be able to provide the necessary energy for the outflow. We computed the jet power (Table \ref{result_sigma}) using the \cite{Birzan 2008} relation
%\begin{equation}
 %   logP_{jet}=(0.35 \pm 0.7)logP_{1400}+(1.85 \pm 0.10)
 %   \label{birzaan}
%\end{equation}

%where $P_{jet}$ is in unit of $10^{42}$erg s$^{-1}$ and $P_{1400}$ is in unit of $10^{24}WHz^{-1}$.   %Finire di fare il conto considerando la polvere. 

\section{Discussion}
\label{discussion}

\subsection{Comparison with ionized outflows in AGN in the literature}
\begin{figure*}[t!]
\centering
    \includegraphics[width = 0.7 \textwidth]{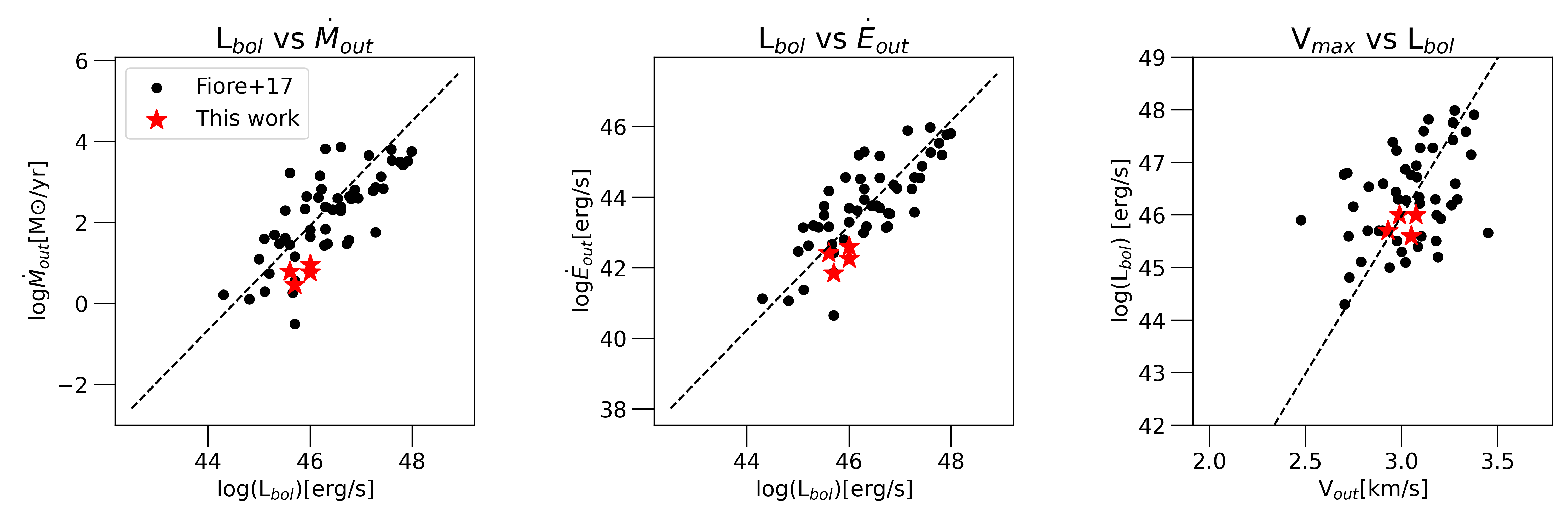}
    \caption{Left: Mass outflow rate as a function of the AGN bolometric luminosity. The black dots mark the ionized outflow measurements from \cite{Fiore2017}, and red stars show our estimate for the four galaxies of our sample. The dashed black line is the best-fit correlation from \cite{Fiore2017} for their ionized outflow sample. Middle: Kinetic power as a function of the AGN bolometric luminosity. Right: AGN bolometric luminosity as a function of the maximum outflow velocity.}
    \label{Fiore+17}
\end{figure*}
We compared our results with other works in the literature.
\cite{Fiore2017} studied the dependence of the properties of outflows on the AGN bolometric luminosity in a sample covering $L_\mathrm{bol,AGN}$ between $ \sim 2\times 10^{44}$ erg s$^{-1}$ and $\sim 10^{48}$ erg s$^{-1}$. For the targets presented in this work, we adopted the AGN bolometric luminosities calculated by \cite{Harrison2014} through mid- to far-infrared Spectral Energy Distribution (SED) fitting to archival photometric data.\\
The left and middle panels of Fig. \ref{Fiore+17} show the mass outflow rates and the kinetic powers as a function of $L_\mathrm{bol,AGN}$. The black dots represent the sample of \cite{Fiore2017}, who found a good correlation with $L_\mathrm{AGN}$  with log-linear slopes of 1.29 $\pm$ 0.38 and 1.50 $\pm$ 0.34 for the two above quantities, respectively (dashed black lines).
Although \cite{Fiore2017} used a lower value for the electron density (200 cm$^{-3}$) for all galaxies in their sample compared to the densities inferred in this work from [SII] (Table \ref{result_fiore}) and they assumed a decreasing mass outflow rate with time (C = 3), we found values for the mass outflow rates and the kinetic power (red stars in Fig. \ref{Fiore+17})  %$\sim10^{0.5-1.5}M_{\odot}yr^{-1}$,
that agree with those found for type 2 AGN with a similar bolometric luminosity (\citealt{Fiore2017}). 
In Fig. \ref{Fiore+17} we show the relation between AGN bolometric luminosity and the wind velocity, v$_\mathrm{out}$, defined as in Eq. \ref{vmax}. The results for the AGN in our sample and those in \ref{Fiore+17} also agree well in this case.   

\subsection{Enhancement of the velocity dispersion perpendicular to the radio jet}
\begin{figure}[t!]
\centering
    \includegraphics[width = 0.5 \textwidth]{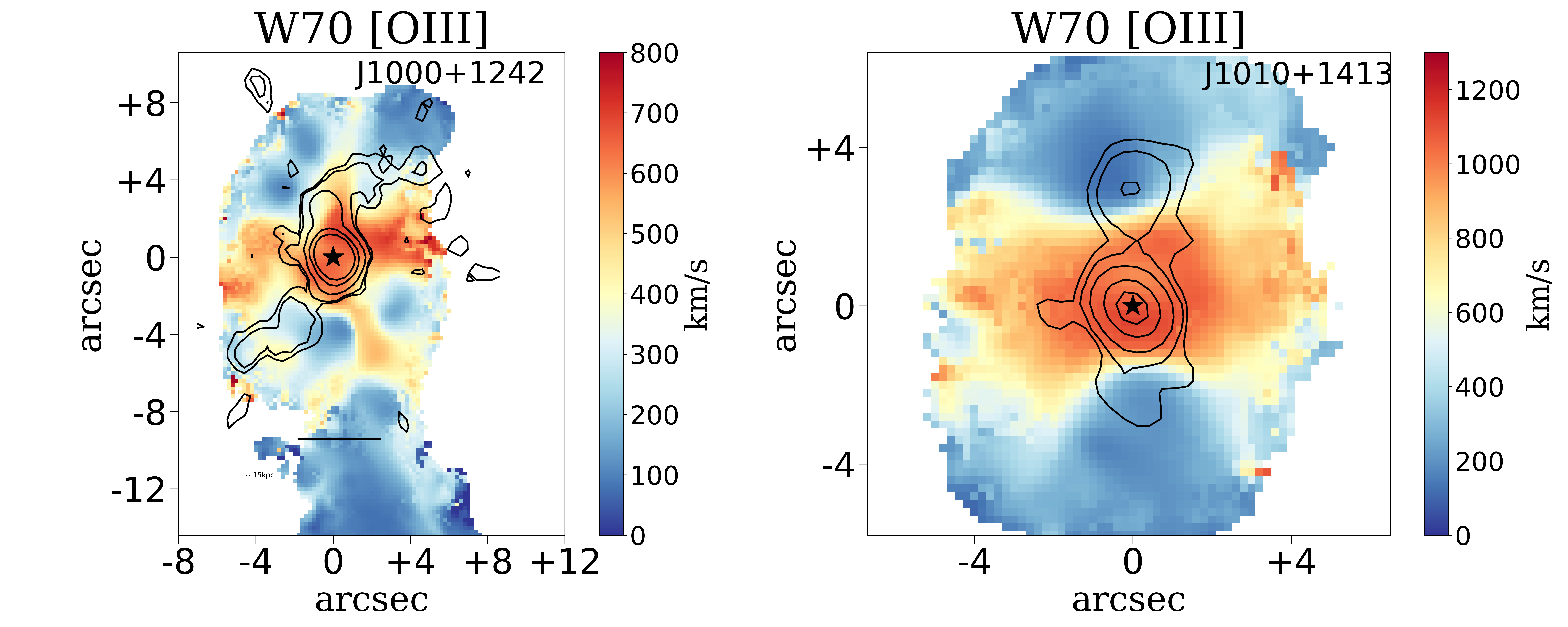}
    \caption{Enhancement of [OIII] W70 perpendicular to the radio jet. The black contours represent the VLA low-resolution ($\sim$1 arcsec beam) radio emission at 6 GHz. J1100+0864 and J1010+0612 are not shown in this plot because the phenomenon studied is not evident due to the poorly resolved radio emission.}
    \label{sigma_radio}
\end{figure}

\begin{figure*}[h!]
    \includegraphics[width = 1 \textwidth]{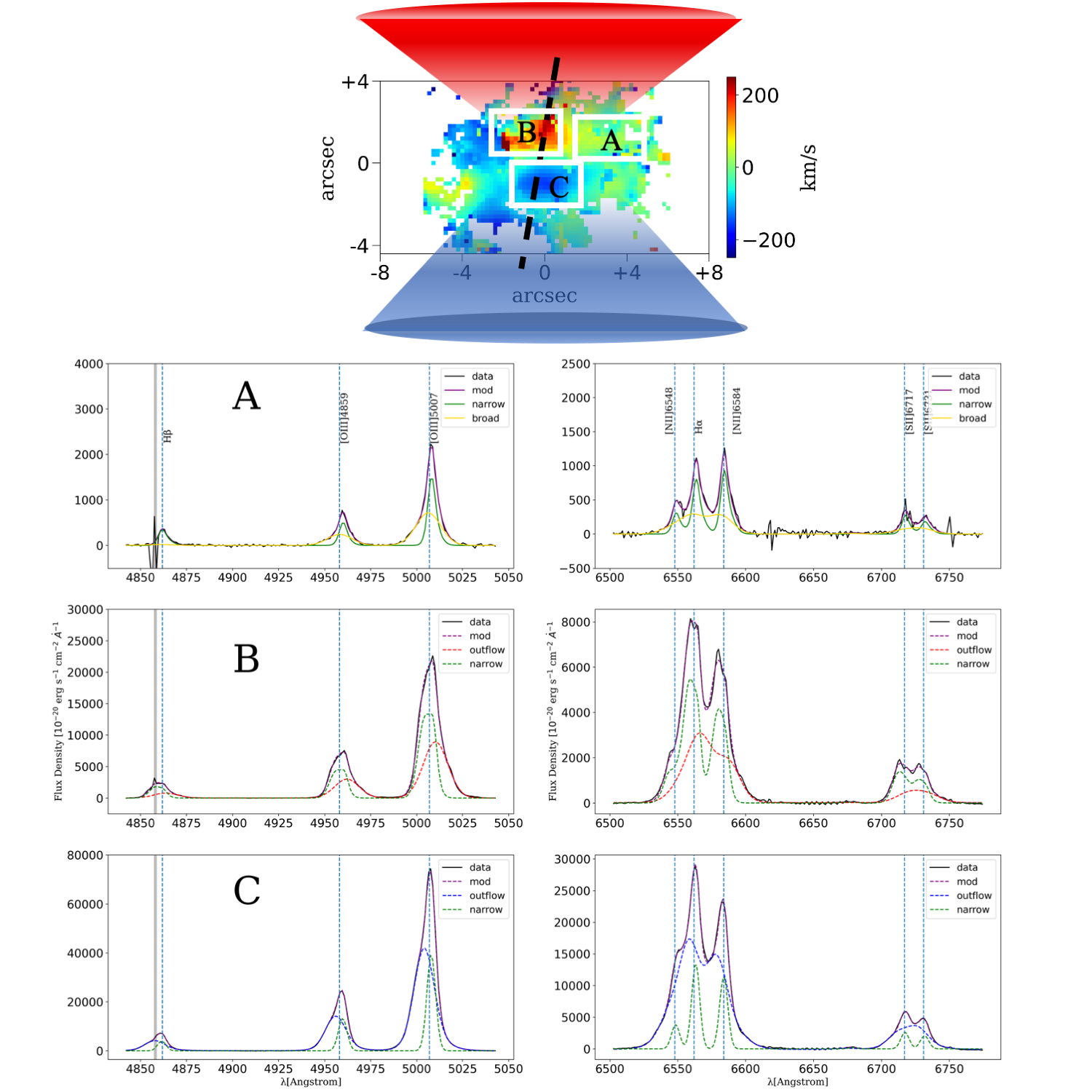}
    \caption{Upper panel: Velocity of the broad component V$_\mathrm{[OIII], BR}$ in J1000+1242, also reported in Fig. \ref{MOM012_J1000-1242}, zoomed in the central region. The axis of the outflowing bicone (traced by the red- and blueshifted motions in regions B and C, respectively) is parallel to that of the radio jet and ionization cones, while turbulent gas is located perpendicular to this (region A). The dashed black lines indicate the ionization cone in which the outflow is detected. The dashed black line indicates the direction of the radio jet. Bottom panels: Integrated spectra extracted from the white rectangles in the corresponding subregions (A, B, and C) in the upper panel. Left panels: H$\beta$, [OIII] doublet. Right panels: H$\alpha$ and the [NII] and [SII] doublets. The data are shown in black, the total best-fitting model is shown in purple, the narrow component is plotted in green, the broad component in region A tracing turbulent gas is shown in yellow, the redshifted outflow is shown in red, and the blueshifted outflow is shown in blue. The shaded gray area indicates the region that was excluded from the fit because the sky subtraction was too poor.}
    \label{sigma_explanation}
\end{figure*}
By comparing the VLA radio maps at low resolution with the maps obtained from our analysis, we note a strongly enhanced gas velocity dispersion in the direction perpendicular to the jet in J1010+1413 and J1010+1242 (Fig. \ref{sigma_radio}). The phenomenon is unclear in J1100+0846, and it is not observed in J1010+0612. In these cases, an enhanced velocity dispersion is detected in the central regions (Figs. \ref{MOM012_J1010-0612} and \ref{MOM012_J1100-0846}), but the radio jet is poorly resolved or unresolved in the data, and thus, it is not straightforward to define a jet direction that can be compared with the high-velocity dispersion region.

We explore the possible origin for this phenomenon in the rest of this section.

\subsubsection{Equatorial outflows}
A possible explanation for the enhancement of the emission line width in the direction perpendicular to the radio jets could be an equatorial outflow \citep[e.g][]{Couto2014,Riffel2015,Freitas2018,Shimizu2019,Ruschel-Dutra+21}, which was predicted by some models to originate from the BH accretion disk. However, we observed significantly blue- and redshifted emission at each side of the nucleus, respectively, that we identified as high-velocity outflows, solely in the direction of the radio jet and the ionization cone.  No net blue- or redshift of the broad component is detected instead in the elongated high-velocity dispersion region. Although no velocity shift is expected from a perfectly radial outflow without extinction, ionized outflows in QSOs commonly exhibit blueshifted wings, particularly in dusty objects, because the receding part of the outflow is located behind the host (\citealt{Bae2014,Carniani2015,Balmaverde2016,Woo2016,Woo2017,Fiore2017}). Moreover, the fact that no clear velocity gradients (blueshifted on one side, redshifted on the other side) in these high-velocity dispersion regions have been observed in other works in the literature \citep[e.g.][]{Venturi2021,Ruschel-Dutra+21}, suggests that the enhancement of the velocity dispersion in the direction perpendicular to the jet and ionization cone is unlikely to be due to an equatorial outflow.

An alternative explanation is that the velocity dispersion enhancement could be the result of beam smearing, which could lead to the blending of spectral line profiles at different velocities along the line of sight, resulting in an increase in the velocity dispersion at the systemic velocity of the galaxy (\citealt{Durre2019,Shin2019}). We exclude the possibility of beam smearing because the scale on which we observe this line-width enhancement is larger than the spatial resolution of the observation.   %We are verifying with the 3D AGN outflow model \MOKA, as presented in Marconcini et al. 2023, if with particular geometric and kinematic configuration of the outflow, it is possible to explain, at least partially, the enhancement of the velocity dispersion perpendicular to the direction of the jet.   %A qualitative illustration is depicted in fig.{} which shows a possible scenario to explain the features of the velocity dispersion encountered in J1010+1413. We made use of the 3D AGN outflow model MOKA 3D, as presented in Marconcini et al. 2023 with the assumption of a biconical outflow to qualitatively recreate the observed situation and we were able to reproduce  

\subsubsection{Turbulence injected by the jet in the ISM}
We conclude that the most plausible explanation for the observed phenomenon is interaction between the radio jet and the ISM of the host galaxy through the injection of turbulent energy in the direction perpendicular to the jet, consistent with \cite{Venturi2021}. As discussed in that work, the increased velocity dispersion in an extended region perpendicular to the jet and ionization cone axis (along which high-velocity outflows are present) is only observed in galaxies hosting a radio jet, even when the radio power is low (<10$^{44}$ erg/s). %Another possible explanation could be the interaction of the radio jet with the galaxy disc that releases energy and gives rise, most likely through shock, to highly turbulent motions in the perpendicular direction. (\cite{Venturi 2021}) 

Fig. \ref{sigma_explanation} summarizes this scenario. The map at the top represents the velocity of the [OIII] broad component (in J1000+1242) that we decoupled from the narrow component, as discussed in Sect. \ref{Analysis}. We identify two regions with different properties. The first region lies parallel to the radio jet and to the ionization cones (labeled B and C),and  the second region lies perpendicular to the radio jet (labeled A). The former presents an [OIII] velocity gradient of $\sim$ 400 km/s, tracing an outflow, while the latter exhibits an [OIII] velocity close to the systemic velocity of the galaxy. Three representative spectra of the regions are shown in the figure. In the spectrum extracted from the southern outflowing cone (C), we observe a blue shifted wing (blue line) that represents the approaching outflow, and in the spectrum extracted from the northern outflowing cone (B), the broad component is redshifted. The spectrum extracted from the region perpendicular to the jet (A) instead presents a fairly symmetric profile with a high velocity dispersion that according to the interpretation discussed above is due to the turbulence injected by the jet in the ISM.\\ %However, we have a big limitation: from the observations it is not possible to establish if this perturbed gas is located on the galaxy disc or whether it is a galactic halo gas out of the galaxy disc.\\
\cite{Venturi2021} reported that this phenomenon occurs exclusively in AGN hosting small-scale ($\sim$1 kpc) low-power ($\lesssim 10^{44}$ erg s$^{-1}$) radio jets with low inclinations ($\lesssim$45$^\circ$) with respect to the galaxy disk (\citealp{Querejeta}). To our knowledge, no line-width enhancement perpendicular to the jet has been detected in galaxies whose radio jets are directed perpendicular to the galaxy disk. At most, an increased line width has been found on very small scales ($\sim$20 - 40 pc) around the jet base. At variance with \cite{Venturi2021}, the angular resolution of our observations does not allow for the determination of the jet inclination with respect to the galaxy disk. Although the inclination of the jet cannot be quantified, the cospatiality of radio hotspots with the most intense emission in ionized gas suggests that the jet and the ISM interact strongly. These aspects strongly suggest that the origin of the enhanced line width perpendicular to the radio jets could be the interaction between the jet and the ISM in the galaxy disk. In support of this scenario, the most recent hydrodynamic simulations of jet-ISM interaction show that when the inclination angle of the radio jet to the galactic disk is high, the radio jet can easily escape, without significant interaction with the disk and without widely broadening the line profile. When the jet is instead coplanar or at low inclinations angles to the disk, the jet remains trapped in the disk and strongly and extensively perturbs the ISM (\citealt{Mukherjee2016,Mukherjee2018,Mukherjee2018a,Meenakshi2022}). In the two targets that clearly exhibit turbulent motions perpendicular to the jet (J1100+0864 and J1010+0612), the morphology of the host galaxy is unclear due to the disturbances induced by the mergers. As a result,  it is not straightforward
 to establish a correlation between this phenomenon and the inclination of the jet with respect to the galaxy disk in our sample. \\
Comparing the the Seyfert galaxies presented by \cite{Venturi2021} %for Seyfert galaxies with luminosity of the AGN $L_{AGN}<10^{44}erg \ s^{-1}$ 
and those analyzed in this work, 
%with $L_{AGN}>10^{45}erg \ s^{-1}$ 
we note that in the latter, the enhanced line velocity widths perpendicular to radio jets are detected over a much larger scale, $\sim$ 10 kpc. The reason probably is that the power of the radio jets in the galaxies of our sample is higher by almost an order of magnitude than in the case of the Seyferts in their work, and for this reason, they are able to induce this phenomenon on a larger scale. 

\begin{figure*}[t!]
\centering
    \includegraphics[width = 1 \textwidth]{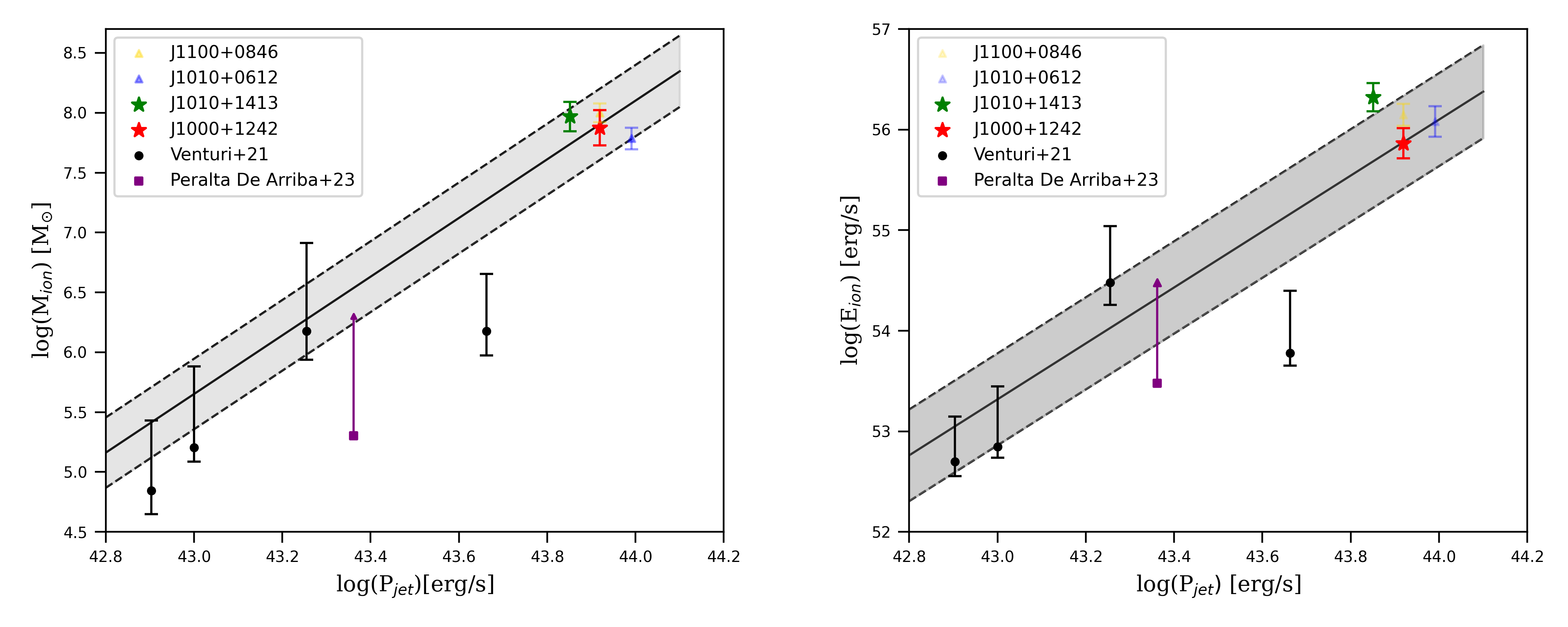}
    \caption{Left: Mass of the gas with a higher velocity dispersion ([OIII] W70 > 300 km/s) as a function of the jet kinetic power. The black dots mark the ionized mass measurements from \cite{Venturi2021} for local Seyfert galaxies, while the colored symbols are our estimates for the ionized mass of the ionized gas in our sample. The stars are used for targets in which the higher velocity dispersion perpendicular to the jet is clearly observed, and the less transparent triangles are used for target in which this phenomenon is unclear, as for J1100+0864, or in which it is not observed, such as J1010+0612.  The quantities have been calculated consistently with  \cite{Venturi2021}. The purple squares mark the measurements from \cite{Peralta2023}, which represent a lower limit of our estimations. Right: Kinetic energy of the high-velocity dispersion gas as a function of jet kinetic power. }
    \label{relation}
\end{figure*}
In order to test whether the described scenario is feasible from the energetic point of view, that is, whether the jets are sufficiently powerful to drive the observed gas perturbations, we estimated the mass of the ionized gas $M_\mathrm{ion}$ that undergoes  this phenomenon and compared it with the jet power. For a direct comparison with the results obtained by \cite{Venturi2021} at lower jet powers (<10$^{44}$ erg/s), we computed the kinetic power of the jets through the following relation from \cite{Birzan2008}:
\begin{equation}
    \log P_\mathrm{jet}=(0.35 \pm 0.7)\,\log P_{1400}+(1.85 \pm 0.10)
    \label{birzaan},
\end{equation}
where $P_\mathrm{jet}$ is the jet kinetic power (in units of $10^{42}$ erg s$^{-1}$), and $P_{1400}$ is its radio luminosity at 1400 MHz (in units of $10^{24}$ W Hz$^{-1}$), taken from \cite{Jarvis2019}.
In order to investigate a possible systematic effect in the estimation of $P_\mathrm{jet}$, we used the relation in \cite{Heckman2014}:
\begin{equation}
    P_\mathrm{jet}= 4 \times 10^{35} \left(f_{W}\right)^{3/2}\left(10 \times P_{1400}\right)^{0.86} ~\mathrm{W}
    \label{heckman},
\end{equation}
where $f_{W}$ is a factor that takes all the uncertainties into account and has a value $\sim$20 (\citealt{Blundell2000}). We find values of $P_\mathrm{jet}$ that are consistent with those found with Eq.\ref{birzaan}, except for J1010-1413, which is slightly beyond the errors. The values are reported in Table \ref{result_sigma}. 

To compute the mass, we first spatially integrated the H$\alpha$ flux from the MUSE maps from regions with [OIII] W70 > 300 km/s (and S/N > 3) and calculated its luminosity. Then, we estimated the mass of the ionized gas through Eq. \ref{Cresci 2017_f}, considering all the emission of H$\alpha$ (not only the broad components) and computing the electron density from the ratio of the total emissions of the [SII] doublet lines, [SII]$\lambda$6717/[SII]$\lambda$6731. % assuming a simplified model by \cite{Genzel 2011} for a case B recombination (\cite{Ost 1989}) of fully ionized gas made up by 90\% of H atoms and 10\% of He atoms with T$\sim$10$^{4}$K.\\ 
By doing so, we included in the turbulent gas all the gas that contributes to the total line emission, which might also include spaxels containing emission from both outflows and disk. Although a clear spatial comparison of the enhanced velocity dispersions with the radio jets is not possible for J1010+0612 and J1100+0846, we computed the turbulent gas mass for these objects as well. They are reported in the plots with fainter markers.
\begin{table*}[h!]\small 
\centering
\caption{Properties of the high-velocity ionized outflow. (1) AGN bolometric luminosity from \cite{Harrison2014}. (2) Total mass in outflow. (3) Outflow electron density estimated from the [SII] doublet ratio. (4) Outflow radius. (5) Maximum outflow velocity, v$_\mathrm{max}$, estimated as in \cite{Fiore2017}. (6) Mass outflow rate. (7) Outflow kinetic power. (8) Total energy in outflow. (9) Coupling efficiency, $\dot{E}_\mathrm{out}/L_\mathrm{AGN}$. The second line for J1010+1413 reports  the outflow properties obtained when considering that all the line emission comes from outflowing gas (see text in Sect. \ref{Results}). 
} 
\begin{tabular}{cccccccccc}
\hline 
\hline
Name & log $L_\mathrm{AGN}^{(1)}$ & $M_\mathrm{out}^{(2)}$ & $n_\mathrm{e,out}^{(3)}$ & $R_\mathrm{out}^{(4)}$ & $v_\mathrm{out}^{(5)}$ & $\dot M_\mathrm{out}^{(6)}$ & $\dot {E}_\mathrm{out}^{(7)}$& $E_\mathrm{out}^{(8)}$ & $\epsilon^{(9)}$ \\
& [erg s$^{-1}$] & [$10^{7}$ M$_\odot$] & [cm$^{-3}$] & [kpc] & [km s$^{-1}$] &  [M$_\odot$ yr$^{-1}$] & [10$^{42}$ erg s$^{-1}$] & [10$^{56}$ erg]&\\
\hline
\\
J1000+1242 & 45.7 & 1.9 &1230 $\pm$ 450 &6& 850 $\pm$ 115 & 2.9 & 0.7 $\pm$ 0.7 & 1.5 $\pm$ 0.5& 4$\times10^{-4}$ \\
\\
%&&&&16.9&19.9&626&4.0$\times10^{-4}$\\
J1010+1413 & 46 &  5.2 & 360 $\pm$ 100 &7 &1190 $\pm$ 100 & 9.1 & 4 $\pm$ 2 & 7.4 $\pm$ 2.4 & 1.2$\times10^{-3}$\\
&&10&&17&&6.7&2.2&12.2&6.7$\times10^{-4}$\\
\\
%&&&&16.5&169&1050&1.7$\times10^{-3}$\\
J1010+0612 & 45.6  & 3.7 & 360 $\pm$ 100&7 & 1123 $\pm$ 155 & 6.2 & 2.6 $\pm$ 2.5 & 4.9 $\pm$ 1.5 &2$\times10^{-3}$\\
\\
%&&&&25.7&44.6&745&1.1$\times10^{-3}$\\
J1100+0846 & 46 & 5.3 &336 $\pm$ 100 & 9 & 977 $\pm$ 120& 6 & 1.8 $\pm$ 2.8 & 5.2 $\pm$ 2.5 & 5.5$\times10^{-4}$\\
%&&&&22.9&26.9&609&2.6$\times10^{-4}$\\

\hline

    \label{result_fiore}

\end{tabular}
\end{table*}
%In this stage, we don't distinguish between different emission, instead, we estimate the overall mass of turbulent gas in the galaxy. This is why we computed the mass for all the targets in our sample even if just two of them show clearly the perpendicular enhancement of velocity dispersion.\\

Fig. \ref{relation} reports the mass of turbulent gas versus the power of the jets. 
By plotting our sample together with the nearby Seyfert galaxies from \cite{Venturi2021,Peralta2023}, we see a correlation between the power of the jet and the turbulent gas mass. The correlation holding over more than one order of magnitude in jet power and $\sim$3 dex in mass and kinetic energy suggests that more powerful jets are able to affect larger quantities of the ISM, and this supports the scenario in which jets cause the enhancement in the emission line width in the perpendicular direction. The best-fit relation is
\begin{equation}
    \log (M_\mathrm{ion})=2.4\log(P_\mathrm{jet})-(98.5\pm0.25),
\end{equation} 
where $M_\mathrm{ion}$ is the mass of ionized gas in solar masses, and $P_\mathrm{jet}$ is the kinetic power of the jet in erg/s.\\ %However, the relation found could be due to the effect of sample selection, in fact the AGNs in our sample, which are characterized by stronger radio jets than those observed in the MAGNUM survey, also exhibit high luminosities, typical of quasars, that may lead to higher computed masses.
To be consistent with \cite{Venturi2021} and \cite{Peralta2023}, we estimated the kinetic energy of the gas in the same region as $E_\mathrm{ion} = M_\mathrm{ion}\sigma_\mathrm{ion}^{2}/2$, where $\sigma_\mathrm{ion}$ is the velocity dispersion of H$\alpha$. We employed the velocity dispersion to calculate the ionized gas energy because it is an indicator of the turbulence produced by the radio jets. 
The right panel in Fig. \ref{relation} shows a correlation between the kinetic energy of the high-velocity dispersion gas and the kinetic power of the radio jets. This indicates that jets can inject energy into the ionized gas and can cause the enhancement of line emission. The best-fit relation is
\begin{equation}
    \log(E_\mathrm{ion})=(2.62\pm 0.01)\log(P_\mathrm{jet})-(59.7\pm0.4).
\end{equation}
The correlations further support the scenario in which the jet injects turbulence in the ISM over a wide range of radio jet powers. In principle, this could be a significant mechanism of feedback because the high turbulence injected by the jet into the ISM could prevent the cooling of the gas and additional star formation (\citealt{Mandal2021}). The results are shown in Table \ref{result_sigma}. \\

\begin{table*}
\caption{Properties of the radio jets and the ionized gas involved in the velocity dispersion enhancement perpendicular to the jets. (1) Radio luminosity from \cite{Jarvis2019}. (2) Kinetic power of the jet estimated from its radio luminosity with Eq. \ref{birzaan} from \cite{Birzan2008}. (3) Kinetic power of the jet estimated from its radio luminosity with Eq. \ref{heckman} from \cite{Heckman2014}. (4)
Electron density from the [SII] doublet ratio. (5) Mass of ionized gas computed with Eq. \ref{Cresci 2017_f}. (6) Kinetic energy of ionized gas, $E_\mathrm{ion} = M_\mathrm{ion}\sigma_\mathrm{ion}^{2}/2$. (7) Jet traveling time. (8) Total kinetic energy of jet $E_\mathrm{jet} = P_\mathrm{jet}t_\mathrm{jet}$.
}
\centering
% \begin{adjustbox}{width= 0.45\textwidth}
\begin{tabular}{ccccccccc}
\hline 
\hline
Name & $\log L_\mathrm{1.4~GHz}^{(1)}$ & $P_\mathrm{jet}^{(2)}$ & 
$P_\mathrm{jet}^{(3)}$ &
$n_{e}^{(4)}$ & 
$M_\mathrm{ion}^{(5)}$ & $E_\mathrm{ion}^{(6)}$ & $t_\mathrm{jet}^{(7)}$ & $E_\mathrm{jet}^{(8)}$\\
&[W Hz$^{-1}$] & [$10^{43}$ erg s$^{-1}]$&
[$10^{43}$ erg s$^{-1}]$& [cm$^{-3}$] & [$10^{7}$ M$_{\odot}$] & [$10^{56}$ erg] & [Myr] & [$10^{56}$ erg] \\
\hline
J1000+1242& 24.2 & 8.3$^{+2.5}_{-2}$ & 7.3 & 510 & 7.5 $\pm$ 2.4 & 0.7 $\pm$ 0.3 & 7.5 & 195\\
J1010+1413& 24 & 7.1$^{+1.8}_{-1.5}$ & 4.9 & 390 &9.3 $\pm$ 2.5 & 2.1 $\pm$ 0.8 & 3.6 & 80\\
J1010+0612& 24.4 & 9.8$^{+3}_{-2.5}$ & 10.9 & 280 & 6.1 $\pm$ 1.4 & 1.2 $\pm$ 0.5 & 0.04 & 1.2 \\
J1100+0846& 24.2 & 8.3$^{+2.5}_{-2}$ & 7.3 & 260 & 10 $\pm$ 2 & 1.4 $\pm$ 0.4 & 0.1 & 2.3\\
%J0958+1439& 23.5 & 4.8 & 0.95$\pm$0.35& 1.07$\pm$0.45 \\
\hline
\end{tabular}
% \end{adjustbox}
\label{result_sigma}
\end{table*}

To infer whether the jet is sufficiently powerful to be a viable mechanism of negative feedback, we compared the total kinetic energy produced by the jet $E_\mathrm{jet}$ in its traveling time $t_\mathrm{jet}$ with the energy computed for the high-velocity dispersion ionized gas $E_\mathrm{ion}$. Assuming that the computed value of $P_\mathrm{jet}$ represents an average value over its traveling time, we can compute $E_\mathrm{jet}$ as $P_\mathrm{jet}t_\mathrm{jet}$ (\citealt{Venturi2021}).
To estimate the jet traveling times, we used Eq. (A1) from \cite{Mukherjee2018a} with the same parameters as employed for IC5063, and we considered the jet radius as the largest linear size computed in \cite{Jarvis2021} (divided by 2). 
The traveling times and the kinetic energies of the jet are listed in Table \ref{result_sigma}. 
Dividing the total energy $E_\mathrm{ion}$  by the kinetic power of the radio jet $E_\mathrm{jet}$, we find values of about 10$^{-3}$--10$^{-2}$ for J1000+1242 and J1010+1413. This indicates that the jet, with a low efficiency of energy transfer, is potentially capable of injecting the required energy into the ISM and might cause the observed phenomenon in radio-quiet and luminous AGN as well. In J1010+0612 and J1100+0846, where the enhanced dispersion velocity is still high in the central region but is not observed in the direction perpendicular to the jet, the energies of the disturbed gas and the radio jets are comparable. For this reason, we cannot exclude for these targets either that the jet contributes to injecting turbulence into the ISM.

\section{Conclusions}
\label{Conclusions}

We presented a study of four luminous type 2 AGN (J1000+1242, J1010+1413, J1010+0612, and J1100+0846) with observations in the optical band obtained with the integral field spectrograph MUSE at VLT. These luminous AGN (L$_\mathrm{bol} \sim$ 10$^{46}$) are classified as radio quiet, although they host a moderate-power ($\sim10^{44}$ erg/s) radio emission. To study the relation between the radio emission and the properties of the ionized gas, we combined our optical analysis with the radio analysis carried out by \cite{Jarvis2019} with VLA and e-MERLIN observations. \\
We investigated the kinematics of the ionized gas through the emission line profiles of [OIII]5007 and H$\alpha$. We identified the outflowing gas by decoupling a narrow emission component tracing the disk or tidal tails and a broad emission component tracing outflows and turbulent gas for all four targets. We detected extended outflows (up to 15 kpc) with high bulk velocities (up to 1000 km/s) that lie in the same direction as the jet and AGN ionization cones in the two targets with extended radio emission (J1000+1242 and J1010+1413). We computed the ionized outflow mass outflow rate and energetics by directly measuring the extinction and electron density in the outflow. We found $M_\mathrm{ion} \sim10^{7-8}M_{\odot} $,   $\dot E_\mathrm{out} \sim$ 10$^{42-43}$ erg~s$^{-1}$ and $\dot M_\mathrm{out}\sim10~M_{\odot}$~yr$^{-1}$.
%The alignment between the outflows and the radio jets as well as the higher power P$_{jet}$ compared to $\dot{E_{out}}$ suggests that the low power jets could represent the dominant mechanism driving the outflows (e.g., \citealt{Nesvadba2008,Alatalo2015,Cresci2015b,Giroletti2017}).\\

The main result of our work is the detection of a strong (up to W70 $\sim$ 1000 km/s) and extended (up to 20 kpc) emission line velocity width enhancement perpendicular to the direction of the radio jets and of the AGN ionization cones in at least two targets of the sample. For the other two galaxies, the radio data are extended on smaller scales than we can probe with our MUSE data. The radio data therefore prevent a reliable assessment of the radio jet direction and interaction with the ISM. Other recent works (e.g., \citealt{Venturi2021,Girdhar2022}) observed the same phenomenon in other AGN %with luminosity up to $L_{AGN}<10^{44}erg \ s^{-1}$ that
hosting jets with a low kinetic power ($P_\mathrm{jet}<10^{44}$erg s$^{-1}$). This was interpreted as due to the interaction between the jet and the ISM in the host galaxy disk. 

Combining the results of our work and those of \cite{Venturi2021}, we found a correlation between the kinetic power of the radio emission and the mass and energy of the ionized gas involved in the phenomenon, in accordance with the possible scenario of an interaction between the radio jet and the ISM. 
We found that the jets are powerful enough to inject the kinetic energy observed in the line width enhancement region into the ISM. This mechanism might represent an important channel of AGN feedback that needs to be taken into account. Moreover, the targets in which we detected this phenomenon have a higher bolometric luminosity and higher radio power emission than most objects in which this phenomenon was observed (but still lower than the powerful jets in radio-loud sources). This suggests that jets with low power might play a significant role in AGN feedback even in radiatively powerful sources (also see \citealt{Girdhar2022}). 
Through spatially resolved S-N BPT diagrams, we found that the main contribution to the ionization arises from the AGN photons, with an enhancement of [NII]/H$\alpha$ in the region with a high velocity dispersion perpendicular to the ionization cones. This may be ascribed to shocks due to the interaction between the jet and the ISM \citep{Mingozzi2023}.\\
However,  to shed light on the effect of the radio jets on the host galaxy and further investigate the incidence and properties of the observed enhanced line widths perpendicular to the radio jets, a larger sample with high-quality integral field spectroscopic data is needed in the optical and  IR bands to assess to which extent the phenomenon affects the molecular phase in addition to the ionized phase.

\section*{Acknowledgements}
This article was produced while attending the PhD program in Space Science and Technology at the University of Trento, Cycle XXXVIII, with the support of a scholarship financed by the Ministerial Decree no. 352 of 9th April 2022, based on the NRRP - funded by the European Union - NextGenerationEU - Mission 4 "Education and Research", Component 1 "Enhancement of the offer of educational services: from nurseries to universities” - Investment 4.1 “Extension of the number of research doctorates and innovative doctorates for public administration and cultural heritage”
GC, AM, GT, FM, FB, EB and GV acknowledge the support of
the INAF Large Grant 2022 "The metal circle: a new sharp view of the baryon cycle up to Cosmic Dawn with the latest generation IFU facilities". GC, AM acknowledge support from PRIN MIUR project “Black Hole winds and the Baryon Life Cycle of Galaxies: the stone guest at the galaxy evolution supper”, contract \# 2017PH3WAT. EDT was supported by the European Research Council (ERC) under grant agreement no. 101040751. SC and GV acknowledge funding from the European Union (ERC, WINGS, 101040227). GV acknowledges support from ANID program FONDECYT Postdoctorado 3200802. CMH acknowledge funding from the United Kingdom Research and Innovation grant (code: MR/V022830/1)

%%%%%%%%%%%%%%%%%%%% REFERENCES %%%%%%%%%%%%%%%%%%
\bibliographystyle{aa}
\bibliography{47436corr} % if your bibtex file is called example.bib

\begin{appendix}
\label{Appendix A}

\section{Stellar kinematic maps}
\begin{figure*}[h!]
\centering
    \includegraphics[width = 1 \textwidth]{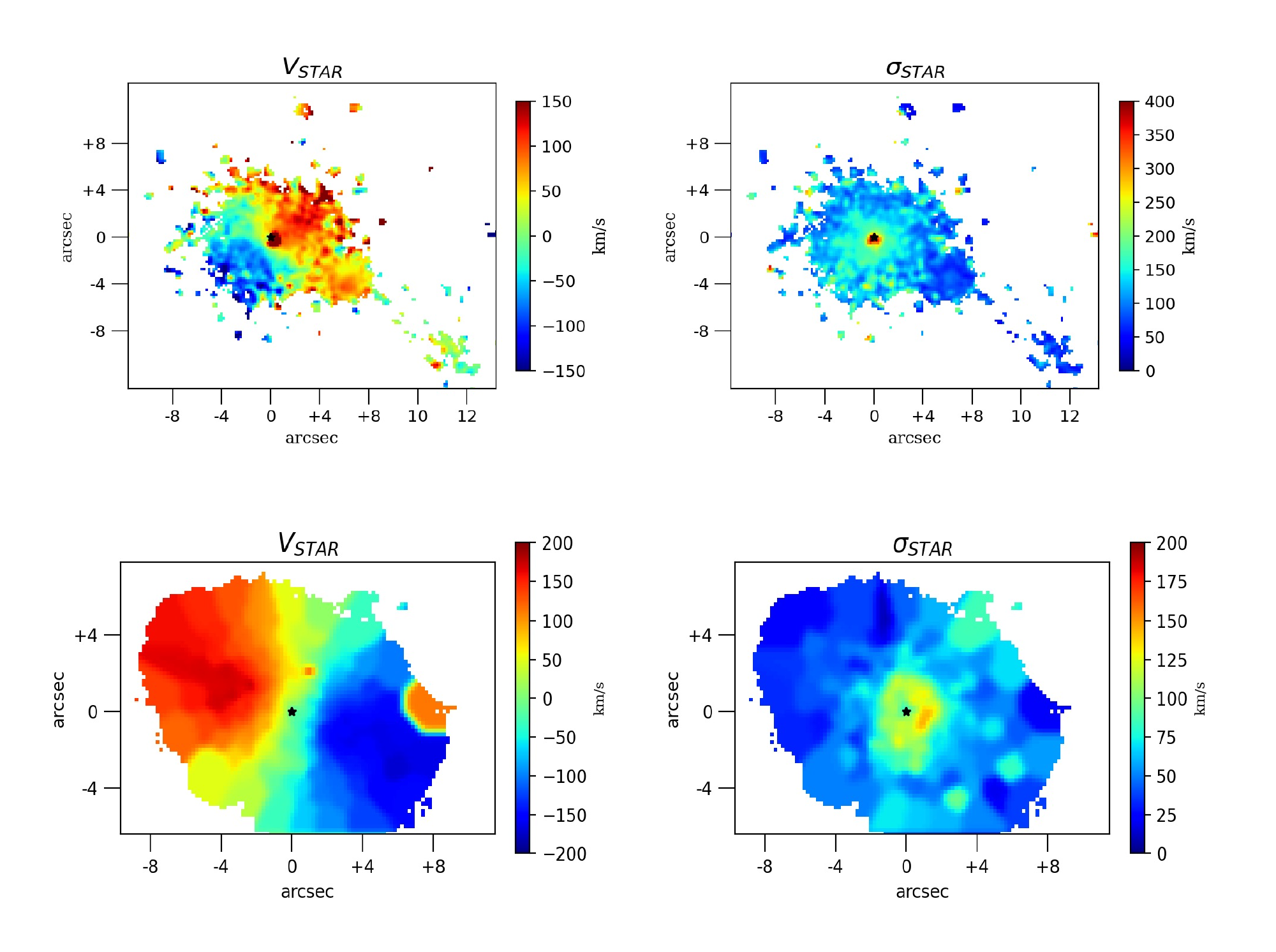}
    \caption{Stellar velocity and velocity dispersion computed with ppxf for J1010+0612 (top panels) and J1100+0846 (bottom panels).
    }
\label{star_kin}

\end{figure*}

In Fig. \ref{star_kin} we present stellar kinematic maps from the modeling with \textsc{PPXF} for J1010+0612 and J1100+0846  to compare them with the narrow component in Fig.\ref{MOM012_J1010-0612} and \ref{MOM012_J1100-0846}. The $\sigma$ was used to decouple the line profile between the narrow and broad component (see text) in these two galaxies. In both galaxies, it is significantly higher in the central region (up to 5 kpc) and decreases in the outer regions (up to 20 kpc). In J1100+0846, the $\sigma$ peaks west of the nucleus. The stellar maps are spatially smoothed using a Gaussian kernel with $\sigma$ = 1 spaxel (0.2$''$) for clarity. 

\end{appendix}

% Don't change these lines
%\bsp	% typesetting comment
\label{lastpage}
\end{document}